\documentclass[prl,aps, twocolumn, superscriptaddress]{revtex4-1}
\usepackage{amsmath,amssymb}
\usepackage{graphicx}
\usepackage{epsfig, color}
\usepackage{wasysym}
\usepackage{amsfonts}
\usepackage{bm}
\usepackage{enumerate}
\usepackage{color}
\usepackage[normalem]{ulem}
\usepackage{mathtools}

\usepackage{latexsym}
\usepackage[breaklinks,colorlinks = true,linkcolor = red,urlcolor=cyan,citecolor=red]{hyperref}

\usepackage[resetlabels]{multibib}
\newcites{supplement}{Supplementary References}

\newcommand{\be}{\begin{eqnarray}}
\newcommand{\ee}{\end{eqnarray}}

\newcommand{\vps}[0]{ \vphantom{*}}
\newcommand{\vpd}[0]{ \vphantom{\dagger}}

\newcommand{\expval}[1]{ \langle #1 \rangle}

\newcommand{\abs}[1]{\left| #1 \right|}

\begin{document}
\title{Quantum Brownian motion in a quasiperiodic potential}
\author{Aaron J. Friedman}
\affiliation{Rudolf Peierls Centre for Theoretical Physics, Clarendon Laboratory, University of Oxford, Oxford, OX1 3PU, UK}
\affiliation{Department of Physics and Astronomy, University of California, Irvine, CA 92697, USA}
\author{Romain Vasseur}
\affiliation{Department of Physics, University of Massachusetts, Amherst, Massachusetts 01003, USA}
\author{Austen Lamacraft}
\affiliation{TCM Group, Cavendish Laboratory, University of Cambridge,
J. J. Thomson Ave., Cambridge CB3 0HE, UK}
\author{S. A. Parameswaran}
\affiliation{Rudolf Peierls Centre for Theoretical Physics, Clarendon Laboratory, University of Oxford, Oxford, OX1 3PU, UK}
\date{\today}

\begin{abstract}
We consider a quantum particle subject to Ohmic dissipation, moving in a bichromatic quasiperiodic potential. In a periodic potential the particle undergoes a zero-temperature localization-delocalization transition as dissipation strength is decreased. We show that the delocalized phase is absent in the quasiperiodic case, even when the deviation from periodicity is infinitesimal.
Using the renormalization group, we determine how the effective localization length 
 depends on the dissipation. We show that {a similar problem can emerge in} the strong-coupling limit of a mobile impurity  moving in a periodic  lattice and immersed in a one-dimensional  quantum gas.
\end{abstract}

\maketitle

\noindent{\textit{Introduction.}---} Localization has been a subject of interest for over half a century, following Anderson's seminal work on electron propagation in disordered media~\cite{PhysRev.109.1492}. Recently, the recognition that the many-body localized (MBL) insulator is a stable {state of matter} with robust non-equilibrium phase structure has sparked renewed interest in the topic~\cite{BAA,2014arXiv1404.0686N,PhysRevB.90.174202,PalHuse,jzi}. Although much of this effort has focused on isolated systems with uncorrelated disorder, two departures from these prevalent paradigms have emerged as significant.  First, studying localization in open quantum systems coupled to an external `bath' is both intrinsically interesting~\cite{MBLBath,GenMBLBath,2014arXiv1402.5971N,hyattSmallBath} and relevant to many experiments~\cite{RevModPhys.80.885,PhysRevLett.120.160404,MBLopenExp,PhysRevX.7.011034}. Second, quasiperiodic systems can also display localization, but unlike their disordered cousins, may be less susceptible to  rare region
effects that disrupt MBL in $d>1$~\cite{AubryAndre,MBLinQP,AandCQPIsing,sggriffiths2,deRoeckImbrie,deRoeckHuveneers,PhysRevLett.119.260401}.
 Quasiperiodic potentials  can be engineered robustly and controllably in cold atom experiments,  either  by superposing two mutually incommensurate optical lattices, or by `cut-and-project' techniques. Experiments have now begun to probe the interplay of localization, interactions, and coupling to a bath in quasiperiodic systems~\cite{Schreiber842,RevModPhys.80.885,SidWeldQuasi,ReyQuasiPeriodicAMO,ReyIncomQPTs,PhysRevLett.120.160404,Rey2007,PhysRevLett.119.260401}.  

Here, we show that the properties of a quasiperiodic system can be altered by coupling to a bath with non-trivial dynamics, even without interactions. As MBL focuses on excited eigenstates and hence high temperature $T$,  baths in that context are approximated as Markovian, i.e. memoryless on long timescales~\cite{GenMBLBath}. In contrast, for $T\rightarrow 0$, the bath autocorrelation time can diverge, so that memory effects become significant.  Such non-Markovian baths can arise naturally from quantum dissipation, induced, e.g. by coupling to a continuum of gapless excitations~\cite{CL1,CL2}. 
   The simplest examples involve dissipative dynamics of a {\it single} quantum degree of freedom~\cite{CL1,CL2,FZ,CallanFreed1,CallanFreed2,WEISS198563,KaneYiQBM}. This can be the position of a particle, but similar models arise more generally in `quantum impurity problems', describing e.g. the phase of a resistively and capacitively shunted Josephson junction, a Kondo spin in a metal, or the scattering phase shift at a quantum point contact or across a mobile impurity in a quantum fluid~\cite{KaneFisherWeakLink,KaneFisherBarriers,MPAFHeavyinLL}. 

Despite their simplicity,  these models can nevertheless exhibit phase transitions, e.g. as a function of dissipation strength~\cite{FZ,WEISS198563,ASLANGUL1,GuineaLattice}. For instance, a particle in a periodic potential can undergo a $T=0$ phase transition as the strength of Ohmic dissipation $\alpha$ is tuned:  for $\alpha> \alpha_c $ the particle is localized in one of the potential minima, while for $\alpha<\alpha_c$ it is delocalized and undergoes quantum Brownian motion over long distances, where $\alpha_c$ is a critical value of dissipation set by the periodicity of the potential~\cite{FZ}. We examine the fate of this $T=0$ transition for {\it quasi}periodic potentials. We show that the delocalized phase present at weak dissipation $\alpha<\alpha_c$ for a single periodic potential~\cite{FZ} is destabilized by an additional periodic perturbation, even when the latter has a {\it higher} critical dissipation strength in isolation. The resulting phase diagram depends on the ratio between the periods of the potentials. In the commensurate case, the delocalized phase survives, but with a lower critical dissipation strength than for either potential in isolation; for the incommensurate (quasiperiodic) case, it is destroyed. Notably, with dissipation the delocalized phase is absent even for {infinitesimally} weak quasiperiodic perturbations, in striking contrast to the dissipationless case~\cite{AubryAndre} where it survives upto a critical value of the quasiperiodicity. Although the problem  {formally maps} to a `double-frequency' boundary sine-Gordon model with no exact solution, we can compute an approximate localization length using renormalization-group (RG) techniques. We showcase this approach for {examples of} commensurate and incommensurate perturbations.

We also find {a surprising application of our analysis to the currently more experimentally realizable setting} of a mobile impurity moving in a periodic lattice in one dimension, immersed in a quantum fluid that it scatters strongly via contact interactions. Here our model describes the dissipative dynamics of the scattering phase across the impurity,  the relevant commensurability is between the gas density and the lattice, and  the transition  corresponds  to a change in the impurity  dispersion (energy-momentum relation $E(P)$), from flat to periodic.

\noindent{\textit{Model.---} } 
We begin by considering a single quantum particle interacting with a bath of harmonic oscillators~\cite{CL1,CL2}. The joint Hamiltonian is
\be H = \label{eq:Hsysbath1} H_0 (q) + \frac{1}{2} \sum_a \frac{p_a^2}{m_a} + m_a \omega_a^2 \left( x_a +\frac{f_a[q] }{m_a \omega_a^2} \right)^2 ,  \ee
where $a$ indexes the oscillators, $q$ is the spatial coordinate of the particle, and $H_0 = p^2/2m + V(q)$, with $V(q)$ a local potential.
We assume linear particle-bath coupling $f[q] = \lambda_a q$, and characterize the bath via its spectral function $J(\omega) = \frac{\pi}{2} \sum_a \frac{\lambda_a^2}{m_a \omega_a} \delta ( \omega - \omega_a)$. 
We restrict to  {Ohmic} dissipation, $J(\omega) = \eta | \omega |$, which in the classical/high-temperature limit yields Brownian motion described by a Langevin equation~\cite{CL1,CL2}. 
Integrating out the bath in the partition function yields an (imaginary-time) effective action for the particle~\cite{FeynmanVernon}, which for Ohmic dissipation and $V=0$ is
\be \label{eq:OGaction} S_0 =  \int\limits_0^{\beta \hbar} d\tau \left[ \frac{m}{2} \dot{q}^2 (\tau) + \frac{\eta}{2 \pi} \int\limits_{-\infty}^{\infty} d\tau' \frac{q(\tau) q(\tau')}{\left(  \tau - \tau' \right)^2}\right]. \ee
We scale out a microscopic length $q_0$ {(this will be set by the potential)} and take $\theta (\tau) = { 2 \pi q (\tau)}/{q_0}$. We identify the characteristic energy scale  $E_0 = {(2 \pi \hbar)^2}/{m q_0^2}$ required to confine the particle to $q_0$, so that $\Lambda= E_0/\hbar$ sets the scale of the bare kinetic energy. 
 {Since this is irrelevant under the RG by power counting (compared to the nonlocal bath contribution) we replace it by a cutoff $\Lambda$ on the bath term}
~\cite{FZ,CallanFreed1,CallanFreed2,KaneYiQBM}  
\be \label{eq:GaussRed} S_0  [\theta ( \omega)]  = \frac{\alpha}{4 \pi} \int\limits_{-\Lambda}^{\Lambda} {\frac{d\omega}{2\pi}}~ | \omega| \left| \theta (\omega) \right|^2. \ee

{Appropriate choices of $V(q)$ realize a number of interesting scenarios.} We will exclusively consider potentials of the form $V(q) = -\sum_{\mu} V_\mu \cos (\lambda_\mu q)$, with one or two $V_\mu$ initially nonzero. In this case, we choose $q_0 = 2\pi/\min[{\lambda_\mu}]$, and rescale parameters to obtain $V[\theta] = \sum_\mu V_\mu \cos(\lambda_\mu \theta)$, where now $\lambda_\mu \geq 1$ and $V_1\neq 0$. We will analyze the phase diagram of $S_0+S_V${, where $S_V = \int d\tau\, V[\theta(\tau)]$,} for different choices of {$\lambda_{\mu}$}.

\noindent{\textit{Single Frequency.---}}  
{We first consider} a single harmonic, i.e. $V_\mu =0 $ for $\mu\neq 1$, corresponding to a particle in a periodic potential~\cite{FZ,ASLANGUL1,CallanFreed1,CallanFreed2,KaneYiQBM,WEISS198563}, with
\be 
\label{eq:LatPot} S_V\!\!\left[ \theta (\tau) \right] = -V_{1}\int d\tau \cos \left[ \theta (\tau) \right], 
\ee
meaning $S_0+S_v$ is a boundary sine-Gordon model. {Therefore, the perturbative effect of the potential to the `free fixed point' \eqref{eq:GaussRed} can be straightforwardly diagnosed using momentum-shell RG}~\cite{FZ,SupMat}, as follows. 
 {First, we} split the fields into `slow' and `fast' modes $\theta(\omega) = \theta_s(\omega)\Theta[\Lambda/b -\omega] + \theta_f(\omega)\Theta[\omega-\Lambda/b]$ where $\Theta$ is the unit step function, and $b=e^{\ell}$. We then integrate out the fast modes, possibly generating new terms, {using a cumulant expansion about the Gaussian $S_0$,} and rescale frequencies via $\omega\mapsto b\omega$ to keep $S_0$ fixed. Finally, we define rescaled fields via ${\theta} \left( \tilde{\omega} \right) = b^{-1} \theta_s (\omega)$. Iterating this transformation, we obtain the RG flow equation for ${V}_1$:
\be 
\label{eq:LatRGflow} \frac{d V_1}{d \ell} = \left( 1 - \frac{1}{\alpha} \right) V_1 + {O}(V_1^3).
 \ee
This shows that the model has a phase transition at $\alpha_c=1${: for} $\alpha<\alpha_c$, $V_1$ flows to zero under the RG (corresponding to the free phase), whereas for $\alpha>\alpha_c$, $V_1$ is relevant and the flow is to strong coupling. In this limit, a variational estimate suggests that the localization length $\xi^*$ diverges as $({\alpha-\alpha_c})^{-1/2}$~\cite{FZ}.  The constancy of $\alpha$ under RG follows from two facts. First, note that $V_1$ is local in time, and coarse-graining preserves locality; in contrast, $S_0$ is nonlocal in time for $T\rightarrow 0$, and so cannot emerge in the perturbative RG. Second, the coefficient of $\theta$ is fixed by translational symmetry, $\theta\rightarrow \theta+2\pi\mathbb{Z}$. Thus, $\alpha$  does not flow~\cite{FZ}. Additionally,  while $V_1$ {itself} does not receive corrections at {second} order in $V_1$,  a $V_2$ term \emph{is} generated at {at $O(V_1^2)$. } However, it is less relevant than $V_1$, which is always the most relevant term generated by the flow to {\it all} orders. 
{(This will no longer be true if a second harmonic $V_{\gamma}$ with $\gamma \not\in \mathbb{Z}$ is included.)}

\noindent{\textit{Generalized RG Flows.---}}  We now study the double-frequency {(bichromatic)} boundary sine-Gordon model, 
\be 
\label{eq:LatPot} S_V \, \left[ \theta (\tau) \right] = -\int d\tau\left\{  V_1\cos \left[ \theta (\tau) \right] +  V_\gamma\cos \left[ \gamma\theta (\tau) \right]\right\}, \,
\ee
where, without loss of generality, we take $\gamma>1$. Observe that with this choice, for $\alpha<1$, both $V_1$ and $V_\gamma$ are irrelevant if considered in isolation.  For $\gamma\in \mathbb{Z}$,  any term generated by the RG has a higher scaling dimension than $V_1$, and is therefore also irrelevant. For $\gamma\not\in \mathbb{Z}$, we must consider the terms generated at second order in the RG equations. Intuitively, this is because `beating' between two cosines can yield a cosine with a shorter wavelength, potentially relevant even when $V_1, V_\gamma$ are not. This picture already signals that rational and irrational $\gamma$ are physically distinct: {in the former case, there are finitely many such beats; in the latter there are infinitely many}. This is a  consequence of the fact that a quasiperiodic potential has no shortest reciprocal lattice vector~\footnote{ Though they are not periodic in space, quasicrystals {\it do} have a regular structure of Bragg peaks.}. 

To study these effects quantitatively, we determine the RG flow equations. We consider  all  wavevectors  generated by the RG, corresponding to the set  $\mathcal{L} = \{\lambda: \lambda = |m+ \gamma n|,\, m,n\in \mathbb{Z}\}$ \footnote{In mathematical terms, $\mathcal{L}$ is a $\mathbb{Z}$-module on $\{1,\gamma\}$.}.  While an explicit derivation of RG equations requires a  tedious (albeit standard) cumulant expansion~\cite{SupMat}, their structure is fixed by the operator product expansion of boundary sine-Gordon theory: 
\be 
\label{eq:DFBSGRGflow} \frac{d V_\lambda}{d \ell} = \left(1 - \frac{\lambda^2}{\alpha} \right) V_\lambda + \sum_{{\lambda',\lambda''}}C^{\lambda'\lambda''}_\lambda V_{\lambda'}V_{\lambda''} + \ldots
\ee
where $C^{\lambda'\lambda''}_\lambda =  \frac{\lambda'\lambda''}{2 \alpha}(\delta_{\lambda, \lambda' + \lambda''} - \delta_{\lambda, \lambda' - \lambda''})$, and `$\ldots$' denotes higher-order terms that we neglect in this perturbative analysis.  
Evidently, this coupled set of equations \eqref{eq:DFBSGRGflow} captures the beat phenomenon described above, since at $O(V^2)$ the RG  generates new terms that are absent at the bare level. These in turn generate other terms as the flow proceeds.  The absence of $\theta\mapsto \theta+2\pi$ symmetry may allow additional terms that in principle could affect the RG flows; however, the set \eqref{eq:DFBSGRGflow} remains valid at a perturbative level, and we proceed assuming their validity.
{To understand their solution}, we consider the scenario where  $V_1 = u_0, V_\gamma = \epsilon u_0$ at the bare level; for a given $\alpha$, the question then is to determine (i) the new critical dissipation strength $\alpha_c' <\alpha_c$; (ii) the RG time $\ell^*(\alpha)$ at which, for $\alpha_c' <\alpha<\alpha_c$ {a} relevant potential generated by these bare values flows to $O(1)${; and (iii) the corresponding localization length associated with this relevant potential}. For $\ell\gtrsim\ell^*$ we enter the strong-coupling regime where our perturbative RG is no longer reliable. Unlike in the conventional single-frequency boundary sine-Gordon problem, there is no exact solution {or duality} to leverage here. Though we have assumed a flow to strong coupling, we cannot rule out the possibility of an intermediate fixed point stabilized by higher-order terms neglected in \eqref{eq:DFBSGRGflow}; this is a question for future analysis.

 \begin{figure}[t]
\includegraphics[width = 0.98\columnwidth]{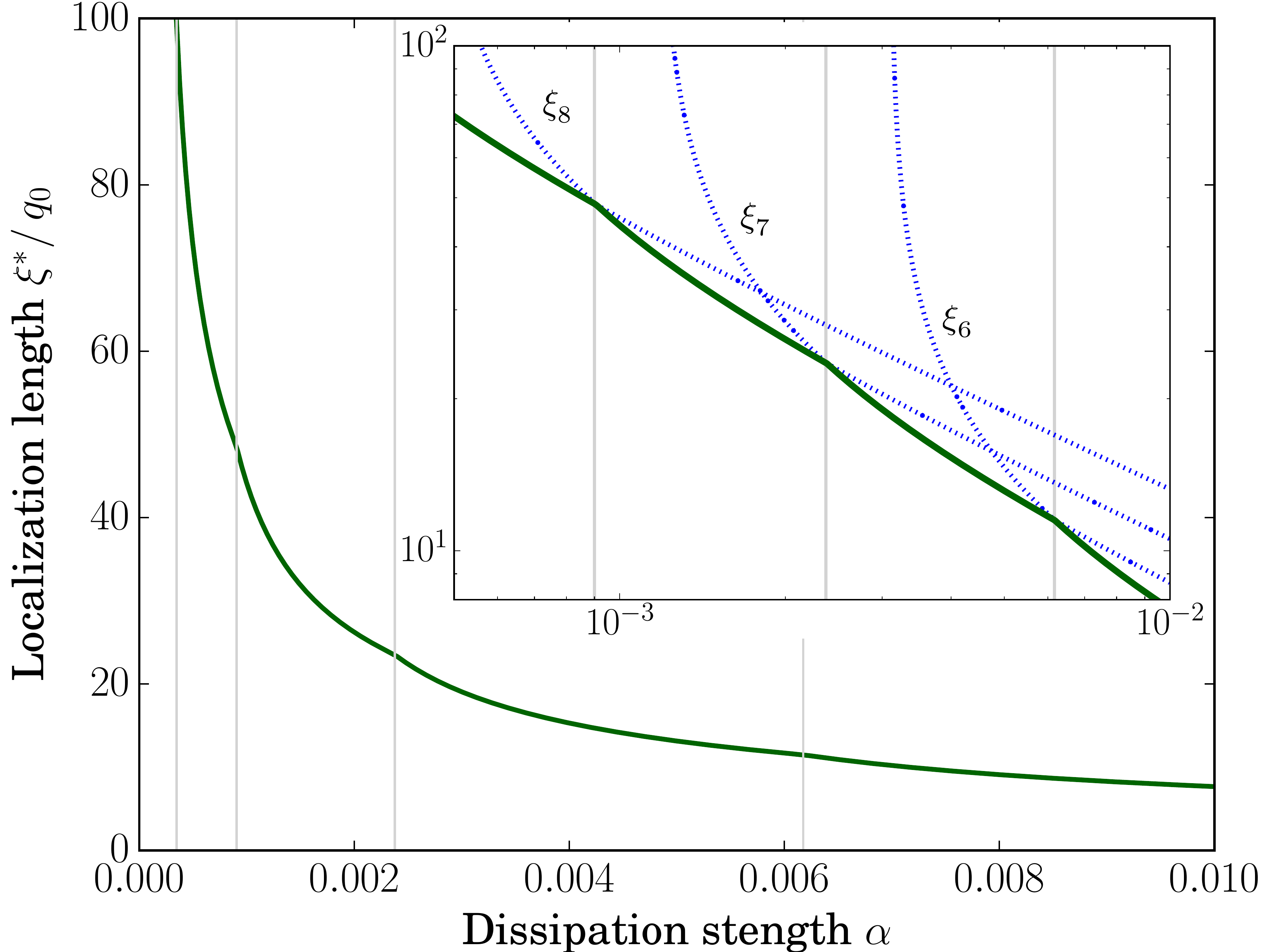} 
\caption{\label{fig:xistar} Localization length $\xi^*$ as a function of dissipation $\alpha$ for quasiperiodic potential with $\gamma =\varphi$ (black). Inset: same plot on log-log scale. As $\alpha$ is decreased, $\xi^{*}$ is a piecewise function that changes non-analytically for $\alpha \sim \alpha_n = \varphi^{-2n}$ between successive $\xi_{n} =\frac{q_0}{2\pi} \sqrt{\frac{2 \ell_n}{\alpha}}$ (see Eq.~\ref{eq:FibLn}). 
}
\end{figure}

{Taking $\gamma  = m/n \geq 1$ to be an irreducible \emph{rational} number, the minimum non-zero wavevector is given by $\lambda_* = 1/n$}, and all  $V_\lambda$ for $\lambda\in \mathcal{L}$  are irrelevant if $\alpha <\alpha_c' \equiv \lambda_*^2$, i.e., the delocalized phase survives, but shrinks in extent. However, for $\alpha_c'<\alpha<1$, the localization is driven by high-order `beats':  bare $V_1, V_\gamma$ are irrelevant, but generate other $V_\lambda$s as they flow to zero; eventually, a relevant term emerges and grows to $O(1)$. The corresponding scale $\ell^*$ controls the crossover to localization: intuitively, it is the scale at which the particle `sees' the potential. To understand this, we consider \eqref{eq:DFBSGRGflow} for a minimal set of $V_{\lambda}$ needed to generate a relevant term. We  ignore second-order terms for each unless they help generate the relevant term, which is justified by numerical iteration of \eqref{eq:DFBSGRGflow}. We then integrate the flows of $V_1 ( \ell )$ and $V_{\gamma} ( \ell)$ directly \cite{SupMat}. For $\gamma=3/2$ and $1/4<\alpha<1$, since a relevant term $(V_{1/2})$ is generated by these two directly, 
 we find it grows to $O(1)$ in an RG `time'
\be \label{eq:Lstar}  \ell^{*} = \frac{\alpha}{\alpha_c - \alpha} \ln \left[ \epsilon u^2_0 \right] + \dots, \ee
where the omitted terms $\dots$ do not involve $ u_0$ or $\epsilon$.
We can extract from this scale a localization length $\xi^* \propto \sqrt{\expval{\theta^2 \left( \tau \right)}}$, where in evaluating the average we only consider the modes between the current RG scale $\Lambda e^{-\ell^{*}}$ and the original cutoff $\Lambda$. We find $\xi^* = \frac{q_0}{2\pi} \sqrt{\frac{2 \ell^*}{\alpha}} \propto (\alpha - \alpha_c)^{-1/2}$~\cite{SupMat}, which mirrors a variational calculation for the single-harmonic problem~\cite{FZ}. A similar relation for $\ell^*$ may be obtained for generic commensurate $\gamma$, but with the difference that higher powers of $\epsilon$ and $u_0$  appear in the logarithm, corresponding to the fact that the relevant operator emerges at a higher order.

\noindent{\textit{Quasiperiodic Case.---}} We now turn to the quasiperiodic (incommensurate) problem. {For} irrational $\gamma \not\in \mathbb{Q}$,  we see immediately that the minimum non-zero wavevector $\lambda^*$ in $\mathcal{L}$ is ill-defined. Therefore, the critical dissipation strength for localization is zero, so that arbitrarily weak dissipation leads to localization. Intuitively, for rational $\gamma =m/n$, the combined potential  $V(\theta) = V_1\cos \theta + V_\gamma \cos\gamma \theta$ always contains a periodic set of equally-spaced minima (e.g., at spacing $2\pi n$); if the dissipation is sufficiently weak that coherent tunneling between these minima remains possible, the delocalized phase survives. Conversely, for irrational $\gamma$, $V(\theta)$ hosts no such periodic set of minima  --- indeed, there is no real-space periodicity. Therefore, the coherent tunneling is disrupted on long length scales, so that no matter how small the dissipation, the particle will eventually come to rest in some potential minimum. 

For concreteness, we consider the \emph{Fibonacci potential}, given by $\gamma=\varphi = \frac{1}{2}\left( 1 + \sqrt{5}\right)$, the Golden mean. Within $\mathcal{L}$, we note that the decreasing sequence $\lambda_n \equiv \left( -1 \right)^n \left( F_{n+1} - \varphi F_n \right) = \varphi^{-n} $ -- where $F_n$ is the $n^{\rm th}$ element of the Fibonacci sequence -- goes to zero rapidly as $n \rightarrow \infty$. We will refer to these as Fibonacci wave numbers: taking $\lambda_0 = 1$, $\lambda_1 = \varphi - 1$ is the first new term generated by the RG with a smaller wave number than those present at the bare level, and subsequent $\lambda_n$ are quickly generated by successive RG iterations, $\lambda_{n} = \lambda_{n-2} - \lambda_{n-1}$. Although for a given 
$\alpha$ there exist many arbitrary $\mu_{m,n} = m- \varphi n$ such that $\mu_{m,n}^2 < \alpha$, a smaller Fibonacci wave number will always have been generated earlier in the RG, and thus will have had more time to grow in strength and spawn further $\lambda_n$. Thus, determining the most relevant wave number is simplified relative to a generic irrational $\gamma$ (though by analogy to the Fibonacci case, we conjecture they will be generated by successive `best rational approximants' of 
$\gamma$).

The crossover to localization is controlled by a critical scale $\ell^{*}$, the RG time for \emph{some} relevant term to become $O(1)$. We denote $\lambda_{n^{*}}$  as the first relevant term become $O(1)$ when all $\lambda_n$ are allowed to be non-zero. Each $\lambda^{\vps}_n$ requires RG time $\ell^{\vps}_n$ to grow to $O(1)$, and $\ell^*$ corresponds to the smallest among the $\ell^{\vps}_n$ for a given $\alpha$, where $\ell_n$ is determined by analogy to \eqref{eq:Lstar}
\be \ell^{\vps}_n = \frac{\alpha }{\varphi^{-2n} - \alpha} \ln \left[ V_{\gamma}^{F_n} V_{1}^{F_{n+1}} \right]  \label{eq:FibLn} \, ,   \ee
as may be verified by direct integration~\cite{SupMat}. Omitted from \eqref{eq:FibLn} are sub-leading corrections that vanish in the limit $\alpha \ll 1$  \cite{SupMat}. As $\alpha$ is decreased, $\ell^*$ is set by successive $\ell_{n^*}$ with larger Fibonacci indices: taking $V^{\vps}_1=-\ln V^{\vps}_{\gamma} = 1$, we see that $\xi^* = \frac{q_0}{2\pi} \sqrt{\frac{2 \ell^*}{\alpha}}$ is determined by successive $\ell_n$ in a piecewise manner, with $\ell^*$ changing from $\ell_n$ to $\ell_{n+1}$ at  $\alpha  \sim \varphi^{-2n} = \lambda_{n}^2$. This leads to non-analyticity in $\xi^*$ (Fig.~\ref{fig:xistar}). Although there is always a relevant, localizing potential with wave number $\lambda_{n^{*}}$, it requires increasingly long for this term to be generated, corresponding to $\ell^{*} \to \infty$. Dynamically, it will take increasingly longer for the particle to `feel' the localization. 

\noindent{\textit{Realization via Mobile Impurity.---}} So far, we have assumed that our model directly describes a particle in a quasiperiodic landscape. This can be challenging to engineer and observe in cold-atom simulations. We now discuss an alternative route to the same physics in a mobile impurity problem~\cite{KaneFisherWeakLink,KaneFisherBarriers,MPAFHeavyinLL,PhysRevB.79.241105}. Consider a single mobile impurity, with coordinate $X$ and momentum $P$, in a periodic optical lattice (spacing $a=1$), and immersed in a quantum fluid. Described the latter as a Luttinger liquid with interaction parameter $K$ and velocity $v$, 
\be
H_{g} = \frac{v}{2\pi}\int dx\, \left[ K (\partial_x\theta)^2 + \frac{1}{K}(\partial_x \phi)^2\right]
\ee
with $[\phi(x),\partial_y\theta(y)] = i\pi \delta(x-y)$ captures its dynamics. We assume that the optical lattice is sufficiently strong that the impurity has tight-binding dispersion given by  $H_{i} = -t_i \cos (P)$, and that the particle and the gas interact via contact interactions $H_{\text{int}} = u\rho(X)$, where $\rho(X)$ is the density of the gas. 
The full Hamiltonian is $H = H_{i} +H_{g}  +H_{\text{int}}$. It is convenient to make a unitary transformation ${\mathcal{U}}_X =  e^{i P_{g} X}$ to  the frame co-moving with the impurity, so that $H\mapsto {\mathcal{U}}_X H {\mathcal{U}}_X^{-1}  = H_{g} +  u \rho(0) - t_i\cos(P -  P_{g})$. Since $X$ is now absent from $H$, $P$ is conserved and corresponds to the total momentum. We now take the $u\rightarrow \infty$ limit, corresponding to a strongly-scattering impurity, where the leading term at $O(1/u)$ 
 involves the tunneling of gas particles across the impurity. This yields the Josephson-like term  $H_{r} \approx -t_g\cos(\Theta)$, where $\Theta= \theta(0^+) - \theta(0^-)$ describes the phase shift across the impurity. We may relate $P_{g}$ to  $\Theta$ by using the usual Luttinger liquid relations for the density $\rho = \pi^{-1}\partial_x\phi$ and momentum $\pi_\phi = \partial_x \theta$:
\be
P_{g} = \int_{|x|>\epsilon} \!\!\!\!dx \rho \pi_\phi = \frac{1}{\pi}\int_{|x|>\epsilon}\!\!\!\! dx\, \partial_x\phi \partial_x\theta = -n \Theta,
\ee
where the integral excludes the origin as there is a break in the fluid at the impurity. We have used the mode expansion $\phi(x) = \phi_0 + \pi \frac{N}{L} x  +\tilde{\phi}(x)$, $\theta(x) = \theta_0 + \pi \frac{J}{L} x  +\tilde{\theta}(x)$, where $N, J$ are the total current and total momentum, respectively, and $n=N/L$. Finally, we integrate out the gapless sound modes of $H_{g}$ subject to the boundary condition $\theta(0^+,t) - \theta(0^-,t) = \Theta(t)$; this generates dissipative dynamics for $\Theta$. Working in imaginary time we arrive at the impurity effective action
\be
\!\!S_{{i}} =\!\!\int\!\!d\tau\!\left[t_i\cos(P\!+\!\gamma \Theta) + t_g\!\cos\Theta \right] +\! \frac{\alpha}{4\pi}\!\!\int\!\!d\omega {|\omega|}\!\left|\Theta_\omega\right|^2\!\!\! \,\,\,\,\,\,\,\,\,
\ee
with $\alpha = 1/K$~\footnote{We take $K$ to be the effective value obtained after eliminating derivative terms of the form $\partial_x\Theta$.}, $\gamma=n$; $P\neq0$ does not affect the RG flows, and hence, we see that the impurity is described by the double-frequency sine-Gordon action, with the wavevector of one of the cosines tuned by the gas density. Reinstating the lattice spacing $a$, we see that $\gamma=na$ corresponds to the number of gas atoms in each unit cell of the potential seen by the impurity; evidently, there is no particular restriction to commensurate $\gamma$. In this language, the regime where the cosines are irrelevant corresponds to an impurity that is non-dispersive, i.e. whose energy is independent of $P$, while the one where the cosines are relevant correspond to a dispersive impurity. When the gas density is commensurate with the impurity potential, the impurity is able to move recoillessly between minima while simultaneously allowing an integer number of gas particles to tunnel across it; for sufficiently weak dissipation this `dressed' process continues to show quantum Brownian motion. This effect is absent in the quasiperiodic case, but depending on the scale at which the system is probed, the dispersion will show different periodicity set by the potential that controls $\xi^*$. We defer further investigation of the impurity realization of the quasiperiodic problem to future work.

\noindent{\textit{Discussion.---}} In conclusion, we have shown that a quantum particle moving in a quasiperiodic potential is always localized by a dissipative bath as $T\rightarrow 0$. This is in sharp contrast with the well-known quantum phase transition in the periodic case. We also argued that this physics could be realized in the strong-coupling regime of a mobile impurity in a one-dimensional Fermi gas moving in a periodic lattice. On the formal side, we note that while the infrared behavior of the single-frequency boundary sine-Gordon field theory can be studied using instanton expansions and integrability, much less is known about multi-frequency variants. It would be very interesting --- and of direct relevance to an experimentally accessible regime of mobile impurity problems --- to develop analytic tools to analyze the flow to strong coupling in this theory, and investigate the possibility of a new class of intermediate-coupling fixed points.

\noindent{\textit{Acknowledgments.---}} We thank R.~Nandkishore, S.~Gopalakrishnan, and S.~Vijay for useful discussions, A.~Nahum for discussions and insightful comments on the manuscript, and S.L.~Sondhi for a question that inspired some of this work. AJF is grateful for the hospitality of the University of Colorado, Boulder, where some of this work was completed. This research was supported by the National Science Foundation via Grants DGE-1321846 (Graduate Research Fellowship Program, AJF) and  DMR-1455366 (SAP), the US Department of Energy, Office of Science, Basic Energy Sciences, under Award No. DE-SC0019168 (RV), and EPSRC Grant No. EP/P034616/1 (AL). All the authors acknowledge the hospitality of the Kavli Institute for Theoretical Physics at the University of California, Santa Barbara, which is supported from NSF Grant PHY-1748958, where part of this work was completed.

\bibliography{QBM.bib}

\begin{thebibliography}{44}%
\makeatletter
\providecommand \@ifxundefined [1]{%
 \@ifx{#1\undefined}
}%
\providecommand \@ifnum [1]{%
 \ifnum #1\expandafter \@firstoftwo
 \else \expandafter \@secondoftwo
 \fi
}%
\providecommand \@ifx [1]{%
 \ifx #1\expandafter \@firstoftwo
 \else \expandafter \@secondoftwo
 \fi
}%
\providecommand \natexlab [1]{#1}%
\providecommand \enquote  [1]{``#1''}%
\providecommand \bibnamefont  [1]{#1}%
\providecommand \bibfnamefont [1]{#1}%
\providecommand \citenamefont [1]{#1}%
\providecommand \href@noop [0]{\@secondoftwo}%
\providecommand \href [0]{\begingroup \@sanitize@url \@href}%
\providecommand \@href[1]{\@@startlink{#1}\@@href}%
\providecommand \@@href[1]{\endgroup#1\@@endlink}%
\providecommand \@sanitize@url [0]{\catcode `\\12\catcode `\$12\catcode
  `\&12\catcode `\#12\catcode `\^12\catcode `\_12\catcode `\%12\relax}%
\providecommand \@@startlink[1]{}%
\providecommand \@@endlink[0]{}%
\providecommand \url  [0]{\begingroup\@sanitize@url \@url }%
\providecommand \@url [1]{\endgroup\@href {#1}{\urlprefix }}%
\providecommand \urlprefix  [0]{URL }%
\providecommand \Eprint [0]{\href }%
\providecommand \doibase [0]{http://dx.doi.org/}%
\providecommand \selectlanguage [0]{\@gobble}%
\providecommand \bibinfo  [0]{\@secondoftwo}%
\providecommand \bibfield  [0]{\@secondoftwo}%
\providecommand \translation [1]{[#1]}%
\providecommand \BibitemOpen [0]{}%
\providecommand \bibitemStop [0]{}%
\providecommand \bibitemNoStop [0]{.\EOS\space}%
\providecommand \EOS [0]{\spacefactor3000\relax}%
\providecommand \BibitemShut  [1]{\csname bibitem#1\endcsname}%
\let\auto@bib@innerbib\@empty
\bibitem [{\citenamefont {Anderson}(1958)}]{PhysRev.109.1492}%
  \BibitemOpen
  \bibfield  {author} {\bibinfo {author} {\bibfnamefont {P.~W.}\ \bibnamefont
  {Anderson}},\ }\href {\doibase 10.1103/PhysRev.109.1492} {\bibfield
  {journal} {\bibinfo  {journal} {Phys. Rev.}\ }\textbf {\bibinfo {volume}
  {109}},\ \bibinfo {pages} {1492} (\bibinfo {year} {1958})}\BibitemShut
  {NoStop}%
\bibitem [{\citenamefont {Basko}\ \emph {et~al.}(2006)\citenamefont {Basko},
  \citenamefont {Aleiner},\ and\ \citenamefont {Altshuler}}]{BAA}%
  \BibitemOpen
  \bibfield  {author} {\bibinfo {author} {\bibfnamefont {D.}~\bibnamefont
  {Basko}}, \bibinfo {author} {\bibfnamefont {I.}~\bibnamefont {Aleiner}}, \
  and\ \bibinfo {author} {\bibfnamefont {B.}~\bibnamefont {Altshuler}},\ }\href
  {\doibase http://dx.doi.org/10.1016/j.aop.2005.11.014} {\bibfield  {journal}
  {\bibinfo  {journal} {Annals of Physics}\ }\textbf {\bibinfo {volume}
  {321}},\ \bibinfo {pages} {1126 } (\bibinfo {year} {2006})}\BibitemShut
  {NoStop}%
\bibitem [{\citenamefont {Nandkishore}\ and\ \citenamefont
  {Huse}(2015)}]{2014arXiv1404.0686N}%
  \BibitemOpen
  \bibfield  {author} {\bibinfo {author} {\bibfnamefont {R.}~\bibnamefont
  {Nandkishore}}\ and\ \bibinfo {author} {\bibfnamefont {D.~A.}\ \bibnamefont
  {Huse}},\ }\href {\doibase 10.1146/annurev-conmatphys-031214-014726}
  {\bibfield  {journal} {\bibinfo  {journal} {Annual Review of Condensed Matter
  Physics}\ }\textbf {\bibinfo {volume} {6}},\ \bibinfo {pages} {15} (\bibinfo
  {year} {2015})}\BibitemShut {NoStop}%
\bibitem [{\citenamefont {Huse}\ \emph {et~al.}(2014)\citenamefont {Huse},
  \citenamefont {Nandkishore},\ and\ \citenamefont
  {Oganesyan}}]{PhysRevB.90.174202}%
  \BibitemOpen
  \bibfield  {author} {\bibinfo {author} {\bibfnamefont {D.~A.}\ \bibnamefont
  {Huse}}, \bibinfo {author} {\bibfnamefont {R.}~\bibnamefont {Nandkishore}}, \
  and\ \bibinfo {author} {\bibfnamefont {V.}~\bibnamefont {Oganesyan}},\ }\href
  {\doibase 10.1103/PhysRevB.90.174202} {\bibfield  {journal} {\bibinfo
  {journal} {Phys. Rev. B}\ }\textbf {\bibinfo {volume} {90}},\ \bibinfo
  {pages} {174202} (\bibinfo {year} {2014})}\BibitemShut {NoStop}%
\bibitem [{\citenamefont {Pal}\ and\ \citenamefont {Huse}(2010)}]{PalHuse}%
  \BibitemOpen
  \bibfield  {author} {\bibinfo {author} {\bibfnamefont {A.}~\bibnamefont
  {Pal}}\ and\ \bibinfo {author} {\bibfnamefont {D.~A.}\ \bibnamefont {Huse}},\
  }\href {\doibase 10.1103/PhysRevB.82.174411} {\bibfield  {journal} {\bibinfo
  {journal} {Phys. Rev. B}\ }\textbf {\bibinfo {volume} {82}},\ \bibinfo
  {pages} {174411} (\bibinfo {year} {2010})}\BibitemShut {NoStop}%
\bibitem [{\citenamefont {Imbrie}(2014)}]{jzi}%
  \BibitemOpen
  \bibfield  {author} {\bibinfo {author} {\bibfnamefont {J.~Z.}\ \bibnamefont
  {Imbrie}},\ }\href@noop {} {\bibfield  {journal} {\bibinfo  {journal} {arXiv
  preprint arXiv:1403.7837}\ } (\bibinfo {year} {2014})}\BibitemShut {NoStop}%
\bibitem [{\citenamefont {Nandkishore}\ \emph
  {et~al.}(2014{\natexlab{a}})\citenamefont {Nandkishore}, \citenamefont
  {Gopalakrishnan},\ and\ \citenamefont {Huse}}]{MBLBath}%
  \BibitemOpen
  \bibfield  {author} {\bibinfo {author} {\bibfnamefont {R.}~\bibnamefont
  {Nandkishore}}, \bibinfo {author} {\bibfnamefont {S.}~\bibnamefont
  {Gopalakrishnan}}, \ and\ \bibinfo {author} {\bibfnamefont {D.~A.}\
  \bibnamefont {Huse}},\ }\href {\doibase 10.1103/PhysRevB.90.064203}
  {\bibfield  {journal} {\bibinfo  {journal} {Phys. Rev. B}\ }\textbf {\bibinfo
  {volume} {90}},\ \bibinfo {pages} {064203} (\bibinfo {year}
  {2014}{\natexlab{a}})}\BibitemShut {NoStop}%
\bibitem [{\citenamefont {Nandkishore}\ and\ \citenamefont
  {Gopalakrishnan}()}]{GenMBLBath}%
  \BibitemOpen
  \bibfield  {author} {\bibinfo {author} {\bibfnamefont {R.}~\bibnamefont
  {Nandkishore}}\ and\ \bibinfo {author} {\bibfnamefont {S.}~\bibnamefont
  {Gopalakrishnan}},\ }\href {\doibase 10.1002/andp.201600181} {\bibfield
  {journal} {\bibinfo  {journal} {Annalen der Physik}\ }\textbf {\bibinfo
  {volume} {529}},\ \bibinfo {pages} {1600181}}\BibitemShut {NoStop}%
\bibitem [{\citenamefont {Nandkishore}\ \emph
  {et~al.}(2014{\natexlab{b}})\citenamefont {Nandkishore}, \citenamefont
  {Gopalakrishnan},\ and\ \citenamefont {Huse}}]{2014arXiv1402.5971N}%
  \BibitemOpen
  \bibfield  {author} {\bibinfo {author} {\bibfnamefont {R.}~\bibnamefont
  {Nandkishore}}, \bibinfo {author} {\bibfnamefont {S.}~\bibnamefont
  {Gopalakrishnan}}, \ and\ \bibinfo {author} {\bibfnamefont {D.~A.}\
  \bibnamefont {Huse}},\ }\href {\doibase 10.1103/PhysRevB.90.064203}
  {\bibfield  {journal} {\bibinfo  {journal} {Phys. Rev. B}\ }\textbf {\bibinfo
  {volume} {90}},\ \bibinfo {pages} {064203} (\bibinfo {year}
  {2014}{\natexlab{b}})}\BibitemShut {NoStop}%
\bibitem [{\citenamefont {Hyatt}\ \emph {et~al.}(2017)\citenamefont {Hyatt},
  \citenamefont {Garrison}, \citenamefont {Potter},\ and\ \citenamefont
  {Bauer}}]{hyattSmallBath}%
  \BibitemOpen
  \bibfield  {author} {\bibinfo {author} {\bibfnamefont {K.}~\bibnamefont
  {Hyatt}}, \bibinfo {author} {\bibfnamefont {J.~R.}\ \bibnamefont {Garrison}},
  \bibinfo {author} {\bibfnamefont {A.~C.}\ \bibnamefont {Potter}}, \ and\
  \bibinfo {author} {\bibfnamefont {B.}~\bibnamefont {Bauer}},\ }\href
  {\doibase 10.1103/PhysRevB.95.035132} {\bibfield  {journal} {\bibinfo
  {journal} {Phys. Rev. B}\ }\textbf {\bibinfo {volume} {95}},\ \bibinfo
  {pages} {035132} (\bibinfo {year} {2017})}\BibitemShut {NoStop}%
\bibitem [{\citenamefont {Bloch}\ \emph {et~al.}(2008)\citenamefont {Bloch},
  \citenamefont {Dalibard},\ and\ \citenamefont {Zwerger}}]{RevModPhys.80.885}%
  \BibitemOpen
  \bibfield  {author} {\bibinfo {author} {\bibfnamefont {I.}~\bibnamefont
  {Bloch}}, \bibinfo {author} {\bibfnamefont {J.}~\bibnamefont {Dalibard}}, \
  and\ \bibinfo {author} {\bibfnamefont {W.}~\bibnamefont {Zwerger}},\ }\href
  {\doibase 10.1103/RevModPhys.80.885} {\bibfield  {journal} {\bibinfo
  {journal} {Rev. Mod. Phys.}\ }\textbf {\bibinfo {volume} {80}},\ \bibinfo
  {pages} {885} (\bibinfo {year} {2008})}\BibitemShut {NoStop}%
\bibitem [{\citenamefont {L\"uschen}\ \emph {et~al.}(2018)\citenamefont
  {L\"uschen}, \citenamefont {Scherg}, \citenamefont {Kohlert}, \citenamefont
  {Schreiber}, \citenamefont {Bordia}, \citenamefont {Li}, \citenamefont
  {Das~Sarma},\ and\ \citenamefont {Bloch}}]{PhysRevLett.120.160404}%
  \BibitemOpen
  \bibfield  {author} {\bibinfo {author} {\bibfnamefont {H.~P.}\ \bibnamefont
  {L\"uschen}}, \bibinfo {author} {\bibfnamefont {S.}~\bibnamefont {Scherg}},
  \bibinfo {author} {\bibfnamefont {T.}~\bibnamefont {Kohlert}}, \bibinfo
  {author} {\bibfnamefont {M.}~\bibnamefont {Schreiber}}, \bibinfo {author}
  {\bibfnamefont {P.}~\bibnamefont {Bordia}}, \bibinfo {author} {\bibfnamefont
  {X.}~\bibnamefont {Li}}, \bibinfo {author} {\bibfnamefont {S.}~\bibnamefont
  {Das~Sarma}}, \ and\ \bibinfo {author} {\bibfnamefont {I.}~\bibnamefont
  {Bloch}},\ }\href {\doibase 10.1103/PhysRevLett.120.160404} {\bibfield
  {journal} {\bibinfo  {journal} {Phys. Rev. Lett.}\ }\textbf {\bibinfo
  {volume} {120}},\ \bibinfo {pages} {160404} (\bibinfo {year}
  {2018})}\BibitemShut {NoStop}%
\bibitem [{\citenamefont {L\"uschen}\ \emph
  {et~al.}(2017{\natexlab{a}})\citenamefont {L\"uschen}, \citenamefont
  {Bordia}, \citenamefont {Hodgman}, \citenamefont {Schreiber}, \citenamefont
  {Sarkar}, \citenamefont {Daley}, \citenamefont {Fischer}, \citenamefont
  {Altman}, \citenamefont {Bloch},\ and\ \citenamefont
  {Schneider}}]{MBLopenExp}%
  \BibitemOpen
  \bibfield  {author} {\bibinfo {author} {\bibfnamefont {H.~P.}\ \bibnamefont
  {L\"uschen}}, \bibinfo {author} {\bibfnamefont {P.}~\bibnamefont {Bordia}},
  \bibinfo {author} {\bibfnamefont {S.~S.}\ \bibnamefont {Hodgman}}, \bibinfo
  {author} {\bibfnamefont {M.}~\bibnamefont {Schreiber}}, \bibinfo {author}
  {\bibfnamefont {S.}~\bibnamefont {Sarkar}}, \bibinfo {author} {\bibfnamefont
  {A.~J.}\ \bibnamefont {Daley}}, \bibinfo {author} {\bibfnamefont {M.~H.}\
  \bibnamefont {Fischer}}, \bibinfo {author} {\bibfnamefont {E.}~\bibnamefont
  {Altman}}, \bibinfo {author} {\bibfnamefont {I.}~\bibnamefont {Bloch}}, \
  and\ \bibinfo {author} {\bibfnamefont {U.}~\bibnamefont {Schneider}},\ }\href
  {\doibase 10.1103/PhysRevX.7.011034} {\bibfield  {journal} {\bibinfo
  {journal} {Phys. Rev. X}\ }\textbf {\bibinfo {volume} {7}},\ \bibinfo {pages}
  {011034} (\bibinfo {year} {2017}{\natexlab{a}})}\BibitemShut {NoStop}%
\bibitem [{\citenamefont {L\"uschen}\ \emph
  {et~al.}(2017{\natexlab{b}})\citenamefont {L\"uschen}, \citenamefont
  {Bordia}, \citenamefont {Hodgman}, \citenamefont {Schreiber}, \citenamefont
  {Sarkar}, \citenamefont {Daley}, \citenamefont {Fischer}, \citenamefont
  {Altman}, \citenamefont {Bloch},\ and\ \citenamefont
  {Schneider}}]{PhysRevX.7.011034}%
  \BibitemOpen
  \bibfield  {author} {\bibinfo {author} {\bibfnamefont {H.~P.}\ \bibnamefont
  {L\"uschen}}, \bibinfo {author} {\bibfnamefont {P.}~\bibnamefont {Bordia}},
  \bibinfo {author} {\bibfnamefont {S.~S.}\ \bibnamefont {Hodgman}}, \bibinfo
  {author} {\bibfnamefont {M.}~\bibnamefont {Schreiber}}, \bibinfo {author}
  {\bibfnamefont {S.}~\bibnamefont {Sarkar}}, \bibinfo {author} {\bibfnamefont
  {A.~J.}\ \bibnamefont {Daley}}, \bibinfo {author} {\bibfnamefont {M.~H.}\
  \bibnamefont {Fischer}}, \bibinfo {author} {\bibfnamefont {E.}~\bibnamefont
  {Altman}}, \bibinfo {author} {\bibfnamefont {I.}~\bibnamefont {Bloch}}, \
  and\ \bibinfo {author} {\bibfnamefont {U.}~\bibnamefont {Schneider}},\ }\href
  {\doibase 10.1103/PhysRevX.7.011034} {\bibfield  {journal} {\bibinfo
  {journal} {Phys. Rev. X}\ }\textbf {\bibinfo {volume} {7}},\ \bibinfo {pages}
  {011034} (\bibinfo {year} {2017}{\natexlab{b}})}\BibitemShut {NoStop}%
\bibitem [{\citenamefont {Aubry}\ and\ \citenamefont
  {Andr\'e}(1980)}]{AubryAndre}%
  \BibitemOpen
  \bibfield  {author} {\bibinfo {author} {\bibfnamefont {S.}~\bibnamefont
  {Aubry}}\ and\ \bibinfo {author} {\bibfnamefont {G.}~\bibnamefont
  {Andr\'e}},\ }\href@noop {} {\bibfield  {journal} {\bibinfo  {journal} {Ann.
  Isr. Phy.}\ }\textbf {\bibinfo {volume} {3}},\ \bibinfo {pages} {133}
  (\bibinfo {year} {1980})}\BibitemShut {NoStop}%
\bibitem [{\citenamefont {Iyer}\ \emph {et~al.}(2013)\citenamefont {Iyer},
  \citenamefont {Oganesyan}, \citenamefont {Refael},\ and\ \citenamefont
  {Huse}}]{MBLinQP}%
  \BibitemOpen
  \bibfield  {author} {\bibinfo {author} {\bibfnamefont {S.}~\bibnamefont
  {Iyer}}, \bibinfo {author} {\bibfnamefont {V.}~\bibnamefont {Oganesyan}},
  \bibinfo {author} {\bibfnamefont {G.}~\bibnamefont {Refael}}, \ and\ \bibinfo
  {author} {\bibfnamefont {D.~A.}\ \bibnamefont {Huse}},\ }\href {\doibase
  10.1103/PhysRevB.87.134202} {\bibfield  {journal} {\bibinfo  {journal} {Phys.
  Rev. B}\ }\textbf {\bibinfo {volume} {87}},\ \bibinfo {pages} {134202}
  (\bibinfo {year} {2013})}\BibitemShut {NoStop}%
\bibitem [{\citenamefont {Chandran}\ and\ \citenamefont
  {Laumann}(2017)}]{AandCQPIsing}%
  \BibitemOpen
  \bibfield  {author} {\bibinfo {author} {\bibfnamefont {A.}~\bibnamefont
  {Chandran}}\ and\ \bibinfo {author} {\bibfnamefont {C.~R.}\ \bibnamefont
  {Laumann}},\ }\href {\doibase 10.1103/PhysRevX.7.031061} {\bibfield
  {journal} {\bibinfo  {journal} {Phys. Rev. X}\ }\textbf {\bibinfo {volume}
  {7}},\ \bibinfo {pages} {031061} (\bibinfo {year} {2017})}\BibitemShut
  {NoStop}%
\bibitem [{\citenamefont {Gopalakrishnan}\ \emph {et~al.}(2015)\citenamefont
  {Gopalakrishnan}, \citenamefont {Agarwal}, \citenamefont {Huse},
  \citenamefont {Demler},\ and\ \citenamefont {Knap}}]{sggriffiths2}%
  \BibitemOpen
  \bibfield  {author} {\bibinfo {author} {\bibfnamefont {S.}~\bibnamefont
  {Gopalakrishnan}}, \bibinfo {author} {\bibfnamefont {K.}~\bibnamefont
  {Agarwal}}, \bibinfo {author} {\bibfnamefont {D.~A.}\ \bibnamefont {Huse}},
  \bibinfo {author} {\bibfnamefont {E.}~\bibnamefont {Demler}}, \ and\ \bibinfo
  {author} {\bibfnamefont {M.}~\bibnamefont {Knap}},\ }\href@noop {} {\bibfield
   {journal} {\bibinfo  {journal} {arXiv preprint arXiv:1511.06389}\ }
  (\bibinfo {year} {2015})}\BibitemShut {NoStop}%
\bibitem [{\citenamefont {{De Roeck}}\ and\ \citenamefont
  {{Imbrie}}(2017)}]{deRoeckImbrie}%
  \BibitemOpen
  \bibfield  {author} {\bibinfo {author} {\bibfnamefont {W.}~\bibnamefont {{De
  Roeck}}}\ and\ \bibinfo {author} {\bibfnamefont {J.~Z.}\ \bibnamefont
  {{Imbrie}}},\ }\href@noop {} {\bibfield  {journal} {\bibinfo  {journal}
  {arXiv:1705.00756}\ } (\bibinfo {year} {2017})}\BibitemShut {NoStop}%
\bibitem [{\citenamefont {De~Roeck}\ and\ \citenamefont
  {Huveneers}(2017)}]{deRoeckHuveneers}%
  \BibitemOpen
  \bibfield  {author} {\bibinfo {author} {\bibfnamefont {W.}~\bibnamefont
  {De~Roeck}}\ and\ \bibinfo {author} {\bibfnamefont {F.~m.~c.}\ \bibnamefont
  {Huveneers}},\ }\href {\doibase 10.1103/PhysRevB.95.155129} {\bibfield
  {journal} {\bibinfo  {journal} {Phys. Rev. B}\ }\textbf {\bibinfo {volume}
  {95}},\ \bibinfo {pages} {155129} (\bibinfo {year} {2017})}\BibitemShut
  {NoStop}%
\bibitem [{\citenamefont {L\"uschen}\ \emph
  {et~al.}(2017{\natexlab{c}})\citenamefont {L\"uschen}, \citenamefont
  {Bordia}, \citenamefont {Scherg}, \citenamefont {Alet}, \citenamefont
  {Altman}, \citenamefont {Schneider},\ and\ \citenamefont
  {Bloch}}]{PhysRevLett.119.260401}%
  \BibitemOpen
  \bibfield  {author} {\bibinfo {author} {\bibfnamefont {H.~P.}\ \bibnamefont
  {L\"uschen}}, \bibinfo {author} {\bibfnamefont {P.}~\bibnamefont {Bordia}},
  \bibinfo {author} {\bibfnamefont {S.}~\bibnamefont {Scherg}}, \bibinfo
  {author} {\bibfnamefont {F.}~\bibnamefont {Alet}}, \bibinfo {author}
  {\bibfnamefont {E.}~\bibnamefont {Altman}}, \bibinfo {author} {\bibfnamefont
  {U.}~\bibnamefont {Schneider}}, \ and\ \bibinfo {author} {\bibfnamefont
  {I.}~\bibnamefont {Bloch}},\ }\href {\doibase 10.1103/PhysRevLett.119.260401}
  {\bibfield  {journal} {\bibinfo  {journal} {Phys. Rev. Lett.}\ }\textbf
  {\bibinfo {volume} {119}},\ \bibinfo {pages} {260401} (\bibinfo {year}
  {2017}{\natexlab{c}})}\BibitemShut {NoStop}%
\bibitem [{\citenamefont {Schreiber}\ \emph {et~al.}(2015)\citenamefont
  {Schreiber}, \citenamefont {Hodgman}, \citenamefont {Bordia}, \citenamefont
  {L\"uschen}, \citenamefont {Fischer}, \citenamefont {Vosk}, \citenamefont
  {Altman}, \citenamefont {Schneider},\ and\ \citenamefont
  {Bloch}}]{Schreiber842}%
  \BibitemOpen
  \bibfield  {author} {\bibinfo {author} {\bibfnamefont {M.}~\bibnamefont
  {Schreiber}}, \bibinfo {author} {\bibfnamefont {S.~S.}\ \bibnamefont
  {Hodgman}}, \bibinfo {author} {\bibfnamefont {P.}~\bibnamefont {Bordia}},
  \bibinfo {author} {\bibfnamefont {H.~P.}\ \bibnamefont {L\"uschen}}, \bibinfo
  {author} {\bibfnamefont {M.~H.}\ \bibnamefont {Fischer}}, \bibinfo {author}
  {\bibfnamefont {R.}~\bibnamefont {Vosk}}, \bibinfo {author} {\bibfnamefont
  {E.}~\bibnamefont {Altman}}, \bibinfo {author} {\bibfnamefont
  {U.}~\bibnamefont {Schneider}}, \ and\ \bibinfo {author} {\bibfnamefont
  {I.}~\bibnamefont {Bloch}},\ }\href {\doibase 10.1126/science.aaa7432} {\
  \textbf {\bibinfo {volume} {349}},\ \bibinfo {pages} {842} (\bibinfo {year}
  {2015})}\BibitemShut {NoStop}%
\bibitem [{\citenamefont {Singh}\ \emph {et~al.}(2015)\citenamefont {Singh},
  \citenamefont {Saha}, \citenamefont {Parameswaran},\ and\ \citenamefont
  {Weld}}]{SidWeldQuasi}%
  \BibitemOpen
  \bibfield  {author} {\bibinfo {author} {\bibfnamefont {K.}~\bibnamefont
  {Singh}}, \bibinfo {author} {\bibfnamefont {K.}~\bibnamefont {Saha}},
  \bibinfo {author} {\bibfnamefont {S.~A.}\ \bibnamefont {Parameswaran}}, \
  and\ \bibinfo {author} {\bibfnamefont {D.~M.}\ \bibnamefont {Weld}},\ }\href
  {\doibase 10.1103/PhysRevA.92.063426} {\bibfield  {journal} {\bibinfo
  {journal} {Phys. Rev. A}\ }\textbf {\bibinfo {volume} {92}},\ \bibinfo
  {pages} {063426} (\bibinfo {year} {2015})}\BibitemShut {NoStop}%
\bibitem [{\citenamefont {Li}\ \emph {et~al.}(2010)\citenamefont {Li},
  \citenamefont {Satija}, \citenamefont {Clark},\ and\ \citenamefont
  {Rey}}]{ReyQuasiPeriodicAMO}%
  \BibitemOpen
  \bibfield  {author} {\bibinfo {author} {\bibfnamefont {S.}~\bibnamefont
  {Li}}, \bibinfo {author} {\bibfnamefont {I.~I.}\ \bibnamefont {Satija}},
  \bibinfo {author} {\bibfnamefont {C.~W.}\ \bibnamefont {Clark}}, \ and\
  \bibinfo {author} {\bibfnamefont {A.~M.}\ \bibnamefont {Rey}},\ }\href
  {\doibase 10.1103/PhysRevE.82.016217} {\bibfield  {journal} {\bibinfo
  {journal} {Phys. Rev. E}\ }\textbf {\bibinfo {volume} {82}},\ \bibinfo
  {pages} {016217} (\bibinfo {year} {2010})}\BibitemShut {NoStop}%
\bibitem [{\citenamefont {He}\ \emph {et~al.}(2012)\citenamefont {He},
  \citenamefont {Satija}, \citenamefont {Clark}, \citenamefont {Rey},\ and\
  \citenamefont {Rigol}}]{ReyIncomQPTs}%
  \BibitemOpen
  \bibfield  {author} {\bibinfo {author} {\bibfnamefont {K.}~\bibnamefont
  {He}}, \bibinfo {author} {\bibfnamefont {I.~I.}\ \bibnamefont {Satija}},
  \bibinfo {author} {\bibfnamefont {C.~W.}\ \bibnamefont {Clark}}, \bibinfo
  {author} {\bibfnamefont {A.~M.}\ \bibnamefont {Rey}}, \ and\ \bibinfo
  {author} {\bibfnamefont {M.}~\bibnamefont {Rigol}},\ }\href {\doibase
  10.1103/PhysRevA.85.013617} {\bibfield  {journal} {\bibinfo  {journal} {Phys.
  Rev. A}\ }\textbf {\bibinfo {volume} {85}},\ \bibinfo {pages} {013617}
  (\bibinfo {year} {2012})}\BibitemShut {NoStop}%
\bibitem [{\citenamefont {Rey}\ \emph {et~al.}(2007)\citenamefont {Rey},
  \citenamefont {Satija},\ and\ \citenamefont {Clark}}]{Rey2007}%
  \BibitemOpen
  \bibfield  {author} {\bibinfo {author} {\bibfnamefont {A.~M.}\ \bibnamefont
  {Rey}}, \bibinfo {author} {\bibfnamefont {I.~I.}\ \bibnamefont {Satija}}, \
  and\ \bibinfo {author} {\bibfnamefont {C.~W.}\ \bibnamefont {Clark}},\ }\href
  {\doibase 10.1134/S1054660X07020260} {\bibfield  {journal} {\bibinfo
  {journal} {Laser Physics}\ }\textbf {\bibinfo {volume} {17}},\ \bibinfo
  {pages} {205} (\bibinfo {year} {2007})}\BibitemShut {NoStop}%
\bibitem [{\citenamefont {Caldeira}\ and\ \citenamefont {Leggett}(1981)}]{CL1}%
  \BibitemOpen
  \bibfield  {author} {\bibinfo {author} {\bibfnamefont {A.~O.}\ \bibnamefont
  {Caldeira}}\ and\ \bibinfo {author} {\bibfnamefont {A.~J.}\ \bibnamefont
  {Leggett}},\ }\href {\doibase 10.1103/PhysRevLett.46.211} {\bibfield
  {journal} {\bibinfo  {journal} {Phys. Rev. Lett.}\ }\textbf {\bibinfo
  {volume} {46}},\ \bibinfo {pages} {211} (\bibinfo {year} {1981})}\BibitemShut
  {NoStop}%
\bibitem [{\citenamefont {Caldeira}\ and\ \citenamefont {Leggett}(1983)}]{CL2}%
  \BibitemOpen
  \bibfield  {author} {\bibinfo {author} {\bibfnamefont {A.}~\bibnamefont
  {Caldeira}}\ and\ \bibinfo {author} {\bibfnamefont {A.}~\bibnamefont
  {Leggett}},\ }\href {\doibase https://doi.org/10.1016/0003-4916(83)90202-6}
  {\bibfield  {journal} {\bibinfo  {journal} {Annals of Physics}\ }\textbf
  {\bibinfo {volume} {149}},\ \bibinfo {pages} {374 } (\bibinfo {year}
  {1983})}\BibitemShut {NoStop}%
\bibitem [{\citenamefont {Fisher}\ and\ \citenamefont {Zwerger}(1985)}]{FZ}%
  \BibitemOpen
  \bibfield  {author} {\bibinfo {author} {\bibfnamefont {M.~P.~A.}\
  \bibnamefont {Fisher}}\ and\ \bibinfo {author} {\bibfnamefont
  {W.}~\bibnamefont {Zwerger}},\ }\href {\doibase 10.1103/PhysRevB.32.6190}
  {\bibfield  {journal} {\bibinfo  {journal} {Phys. Rev. B}\ }\textbf {\bibinfo
  {volume} {32}},\ \bibinfo {pages} {6190} (\bibinfo {year}
  {1985})}\BibitemShut {NoStop}%
\bibitem [{\citenamefont {Callan}\ and\ \citenamefont
  {Freed}(1992)}]{CallanFreed1}%
  \BibitemOpen
  \bibfield  {author} {\bibinfo {author} {\bibfnamefont {C.~G.}\ \bibnamefont
  {Callan}}\ and\ \bibinfo {author} {\bibfnamefont {D.}~\bibnamefont {Freed}},\
  }\href {\doibase https://doi.org/10.1016/0550-3213(92)90400-6} {\bibfield
  {journal} {\bibinfo  {journal} {Nuclear Physics B}\ }\textbf {\bibinfo
  {volume} {374}},\ \bibinfo {pages} {543 } (\bibinfo {year}
  {1992})}\BibitemShut {NoStop}%
\bibitem [{\citenamefont {Callan}\ \emph {et~al.}(1993)\citenamefont {Callan},
  \citenamefont {Felce},\ and\ \citenamefont {Freed}}]{CallanFreed2}%
  \BibitemOpen
  \bibfield  {author} {\bibinfo {author} {\bibfnamefont {C.~G.}\ \bibnamefont
  {Callan}}, \bibinfo {author} {\bibfnamefont {A.~G.}\ \bibnamefont {Felce}}, \
  and\ \bibinfo {author} {\bibfnamefont {D.~E.}\ \bibnamefont {Freed}},\ }\href
  {\doibase https://doi.org/10.1016/0550-3213(93)90517-S} {\bibfield  {journal}
  {\bibinfo  {journal} {Nuclear Physics B}\ }\textbf {\bibinfo {volume}
  {392}},\ \bibinfo {pages} {551 } (\bibinfo {year} {1993})}\BibitemShut
  {NoStop}%
\bibitem [{\citenamefont {Weiss}\ and\ \citenamefont
  {Grabert}(1985)}]{WEISS198563}%
  \BibitemOpen
  \bibfield  {author} {\bibinfo {author} {\bibfnamefont {U.}~\bibnamefont
  {Weiss}}\ and\ \bibinfo {author} {\bibfnamefont {H.}~\bibnamefont
  {Grabert}},\ }\href {\doibase https://doi.org/10.1016/0375-9601(85)90517-1}
  {\bibfield  {journal} {\bibinfo  {journal} {Physics Letters A}\ }\textbf
  {\bibinfo {volume} {108}},\ \bibinfo {pages} {63 } (\bibinfo {year}
  {1985})}\BibitemShut {NoStop}%
\bibitem [{\citenamefont {Yi}\ and\ \citenamefont {Kane}(1998)}]{KaneYiQBM}%
  \BibitemOpen
  \bibfield  {author} {\bibinfo {author} {\bibfnamefont {H.}~\bibnamefont
  {Yi}}\ and\ \bibinfo {author} {\bibfnamefont {C.~L.}\ \bibnamefont {Kane}},\
  }\href {\doibase 10.1103/PhysRevB.57.R5579} {\bibfield  {journal} {\bibinfo
  {journal} {Phys. Rev. B}\ }\textbf {\bibinfo {volume} {57}},\ \bibinfo
  {pages} {R5579} (\bibinfo {year} {1998})}\BibitemShut {NoStop}%
\bibitem [{\citenamefont {Kane}\ and\ \citenamefont
  {Fisher}(1992{\natexlab{a}})}]{KaneFisherWeakLink}%
  \BibitemOpen
  \bibfield  {author} {\bibinfo {author} {\bibfnamefont {C.~L.}\ \bibnamefont
  {Kane}}\ and\ \bibinfo {author} {\bibfnamefont {M.~P.~A.}\ \bibnamefont
  {Fisher}},\ }\href {\doibase 10.1103/PhysRevLett.68.1220} {\bibfield
  {journal} {\bibinfo  {journal} {Phys. Rev. Lett.}\ }\textbf {\bibinfo
  {volume} {68}},\ \bibinfo {pages} {1220} (\bibinfo {year}
  {1992}{\natexlab{a}})}\BibitemShut {NoStop}%
\bibitem [{\citenamefont {Kane}\ and\ \citenamefont
  {Fisher}(1992{\natexlab{b}})}]{KaneFisherBarriers}%
  \BibitemOpen
  \bibfield  {author} {\bibinfo {author} {\bibfnamefont {C.~L.}\ \bibnamefont
  {Kane}}\ and\ \bibinfo {author} {\bibfnamefont {M.~P.~A.}\ \bibnamefont
  {Fisher}},\ }\href {\doibase 10.1103/PhysRevB.46.15233} {\bibfield  {journal}
  {\bibinfo  {journal} {Phys. Rev. B}\ }\textbf {\bibinfo {volume} {46}},\
  \bibinfo {pages} {15233} (\bibinfo {year} {1992}{\natexlab{b}})}\BibitemShut
  {NoStop}%
\bibitem [{\citenamefont {Castro~Neto}\ and\ \citenamefont
  {Fisher}(1996)}]{MPAFHeavyinLL}%
  \BibitemOpen
  \bibfield  {author} {\bibinfo {author} {\bibfnamefont {A.~H.}\ \bibnamefont
  {Castro~Neto}}\ and\ \bibinfo {author} {\bibfnamefont {M.~P.~A.}\
  \bibnamefont {Fisher}},\ }\href {\doibase 10.1103/PhysRevB.53.9713}
  {\bibfield  {journal} {\bibinfo  {journal} {Phys. Rev. B}\ }\textbf {\bibinfo
  {volume} {53}},\ \bibinfo {pages} {9713} (\bibinfo {year}
  {1996})}\BibitemShut {NoStop}%
\bibitem [{\citenamefont {Aslangul}\ \emph {et~al.}(1985)\citenamefont
  {Aslangul}, \citenamefont {Pottier},\ and\ \citenamefont
  {Saint-James}}]{ASLANGUL1}%
  \BibitemOpen
  \bibfield  {author} {\bibinfo {author} {\bibfnamefont {C.}~\bibnamefont
  {Aslangul}}, \bibinfo {author} {\bibfnamefont {N.}~\bibnamefont {Pottier}}, \
  and\ \bibinfo {author} {\bibfnamefont {D.}~\bibnamefont {Saint-James}},\
  }\href {\doibase https://doi.org/10.1016/0375-9601(85)90570-5} {\bibfield
  {journal} {\bibinfo  {journal} {Physics Letters A}\ }\textbf {\bibinfo
  {volume} {111}},\ \bibinfo {pages} {175 } (\bibinfo {year}
  {1985})}\BibitemShut {NoStop}%
\bibitem [{\citenamefont {Guinea}\ \emph {et~al.}(1985)\citenamefont {Guinea},
  \citenamefont {Hakim},\ and\ \citenamefont {Muramatsu}}]{GuineaLattice}%
  \BibitemOpen
  \bibfield  {author} {\bibinfo {author} {\bibfnamefont {F.}~\bibnamefont
  {Guinea}}, \bibinfo {author} {\bibfnamefont {V.}~\bibnamefont {Hakim}}, \
  and\ \bibinfo {author} {\bibfnamefont {A.}~\bibnamefont {Muramatsu}},\ }\href
  {\doibase 10.1103/PhysRevLett.54.263} {\bibfield  {journal} {\bibinfo
  {journal} {Phys. Rev. Lett.}\ }\textbf {\bibinfo {volume} {54}},\ \bibinfo
  {pages} {263} (\bibinfo {year} {1985})}\BibitemShut {NoStop}%
\bibitem [{\citenamefont {Feynman}\ and\ \citenamefont
  {Vernon}(1963)}]{FeynmanVernon}%
  \BibitemOpen
  \bibfield  {author} {\bibinfo {author} {\bibfnamefont {R.}~\bibnamefont
  {Feynman}}\ and\ \bibinfo {author} {\bibfnamefont {F.}~\bibnamefont
  {Vernon}},\ }\href {\doibase https://doi.org/10.1016/0003-4916(63)90068-X}
  {\bibfield  {journal} {\bibinfo  {journal} {Annals of Physics}\ }\textbf
  {\bibinfo {volume} {24}},\ \bibinfo {pages} {118 } (\bibinfo {year}
  {1963})}\BibitemShut {NoStop}%
\bibitem [{Sup()}]{SupMat}%
  \BibitemOpen
  \href@noop {} {}\bibinfo {note} {{See supplementary material for details of
  RG procedure and plots of flow of couplings under the RG.}}\BibitemShut
  {Stop}%
\bibitem [{Note1()}]{Note1}%
  \BibitemOpen
  \bibinfo {note} {Though they are not periodic in space, quasicrystals
  {\protect \it do} have a regular structure of Bragg peaks.}\BibitemShut
  {Stop}%
\bibitem [{Note2()}]{Note2}%
  \BibitemOpen
  \bibinfo {note} {In mathematical terms, $\protect \mathcal {L}$ is a
  $\protect \mathbb {Z}$-module on $\protect \{1,\gamma \protect
  \}$.}\BibitemShut {Stop}%
\bibitem [{\citenamefont {Lamacraft}(2009)}]{PhysRevB.79.241105}%
  \BibitemOpen
  \bibfield  {author} {\bibinfo {author} {\bibfnamefont {A.}~\bibnamefont
  {Lamacraft}},\ }\href {\doibase 10.1103/PhysRevB.79.241105} {\bibfield
  {journal} {\bibinfo  {journal} {Phys. Rev. B}\ }\textbf {\bibinfo {volume}
  {79}},\ \bibinfo {pages} {241105} (\bibinfo {year} {2009})}\BibitemShut
  {NoStop}%
\bibitem [{Note3()}]{Note3}%
  \BibitemOpen
  \bibinfo {note} {We take $K$ to be the effective value obtained after
  eliminating derivative terms of the form $\partial _x\Theta $.}\BibitemShut
  {Stop}%
\end{thebibliography}%

\newpage

\setcounter{equation}{0}
\setcounter{figure}{0}
\setcounter{table}{0}
\setcounter{page}{1}
\makeatletter
\renewcommand{\theequation}{S\arabic{equation}}
\renewcommand{\thefigure}{S\arabic{figure}}
\renewcommand{\bibnumfmt}[1]{[S#1]}
\renewcommand{\citenumfont}[1]{S#1}

\begin{appendix}
\onecolumngrid

\begin{center}
\textbf{\large SUPPLEMENTAL MATERIALS}
\end{center}

\section{General details of the model}
\subsection{Integrating out the bath}
Our system consists of a bath of quantum simple harmonic oscillators coupled linearly to a particle's spatial degree of freedom,
\be H = \label{eq:Hsysbath1} H_0 (q) + \frac{1}{2} \sum_a \frac{p_a^2}{m_a} + m_a \omega_a^2 \left( x_a +\frac{f_a[q] }{m_a \omega_a^2} \right)^2 ,  \ee
where $H_0 = p^2/2m + V(q)$, $a$ indexes the oscillators, and the counter-term is implicit. We then calculate the partition function using the Euclidean action in the Matsubara formalism, and exactly integrate out the bath degrees of freedom to obtain an effective action for the particle only. This gives
\be S^{\prime}_{\rm eff} [q] = -\sum_a \left\{ \frac{\lambda^2_a}{4 M_a\omega_a} \int\limits_0^{\beta \hbar} {\rm d} \tau \int\limits_{-\infty}^{\infty} {\rm d} \tau^{\prime} q (\tau) q(\tau^{\prime} ) e^{-\omega_a \left| \tau - \tau^{\prime} \right|} \right\}, \ee
and we identify the bath spectral function
\be \label{eq:Jdef} J \left( \omega \right) = \frac{\pi}{2} \sum_a \frac{\lambda_a^2}{M_a \omega_a} \delta ( \omega - \omega_a ), \ee
which we can constrain to the \emph{Ohmic} form, $J \left( \omega \right) = \eta \left| \omega \right|$ to recover the effective action
\be \label{eq:start} S^{\prime}_{\rm eff} [q]=  -\int\limits_0^{\beta \hbar} {\rm d} \tau \int\limits_{-\infty}^{\infty} {\rm d} \tau^{\prime} q(\tau)  q(\tau^{\prime}) \int\limits_0^{\infty} \frac{{\rm d} \omega}{2 \pi} J (\omega) e^{-\omega \left| \tau - \tau^{\prime} \right|} . \ee

We then define dimensionless quantities: $\theta \left( \tau \right) = \frac{2 \pi}{q_0} q \left( \tau \right)$, $\Lambda = E_0/\hbar$, $E_0 = \left( 2 \pi \hbar \right)^2/ \left( m q_0^2 \right)$, $\alpha = \eta q_0^2 / 2 \pi \hbar$, and Fourier transform to recover
\be \label{eq:act1b} S_0 \left[ \theta \left( \omega \right) \right] = \frac{1}{2 \Lambda} \int\limits_{-\infty}^{\infty} \frac{d\omega}{2\pi} \omega^2 \left| \theta \left( \omega \right) \right|^2 + \frac{\alpha}{4 \pi} \int\limits_{-\infty}^{\infty} \frac{d\omega}{2\pi}  \left| \omega \right| \left| \theta \left( \omega \right) \right|^2 . \ee
If we choose the latter term, which comes from integrating out the bath, as the fixed point action, then we note that the former, kinetic term, will be irrelevant by power-counting. 

\subsection{RG procedure: integrating out fast modes, cumulants}
Recognizing that the kinetic part of \eqref{eq:act1b} is \emph{irrelevant by power-counting}, we can omit this term and restrict the bounds of integration over $\omega$ to $\pm \Lambda$:
\be \label{eq:newS0} S_0 \left[ \theta \left( \omega \right) \right] = \int\limits_{-\Lambda}^{\Lambda} \frac{d\omega}{2\pi} \frac{ \alpha \left| \omega \right|}{4 \pi} \left| \theta \left( \omega \right) \right|^2. \ee
We now \emph{define} the slow and fast modes as $\theta \left( \omega \right)  = \theta_f \left( \omega \right)$ for $\Lambda/b \leq \left| \omega \right| \leq \Lambda$ and  $\theta \left( \omega \right)  = \theta_s \left( \omega \right)$ for $0 \leq \left| \omega \right| \leq \Lambda/b$, where $b \geq 1$ is the parameter for rescaling of frequencies. We will always take the limit $b \rightarrow 1$ in the `momentum shell' procedure, so that only an infinitesimal number of modes are integrated out at each step. In general, the effective theory for the low energy modes is given by the action
\be \label{eq:SEff} \tilde{S} \left[ \theta_s \right] = S_0  \left[ \theta_s \right] + \sum_{\lambda}  \expval{S_{\lambda} \left[ \theta \right] }_{0,f} - \frac{1}{2} \expval{ \left[ \sum_{\lambda} \left(  S_{\lambda} \left[ \theta \right] - \expval{  S_{\lambda} \left[ \theta \right]}_{0,f} \right) \right]^2 }_{0,f} + \dots, \ee
where the $0,f$ subscript on the expectation values indicates that the expectation values apply to the fast fields only, and with respect to the bare action $S_0$. The $\dots$ represent higher cumulant expressions, although we will find these do not contribute to the RG flow.  $S_0  \left[ \theta_s \right] $ is simply the same bare action \eqref{eq:newS0} with the bounds of integration restricted to the `slow modes.'

\subsection{Two point correlation function, $G_{0,f} (\tau, \tau')$}
The two-point correlation function $G$, will be necessary in formulating the second and higher cumulants. Denoting by `$f$' integration over the range $\Lambda/b \leq \abs{\omega} \leq \Lambda$, this is
\begin{align} &G_{0,f} (\tau, \tau') = \left\langle \theta_f (\tau)  \theta_f (\tau') \right\rangle_{0,f} = \int_f \frac{d \omega}{2 \pi} \int_f \frac{d \omega'}{2 \pi} \left\langle \theta_f^{*} (\omega') \theta_f (\omega) e^{i (\omega' \tau' - \omega \tau) } \right\rangle_{0,f} \\
&~~~= \int_f \frac{d \omega}{2 \pi} \int\limits_{\Lambda/b \leq | \omega' | \leq \Lambda} \frac{d \omega'}{2 \pi} e^{i (\omega' \tau' - \omega \tau)} \left( \frac{2 \pi}{\alpha | \omega| } \right) \left( 2\pi \delta(\omega-\omega') \right)
= \frac{2}{\alpha} \int\limits_{\Lambda/b}^{\Lambda}  \frac{d \omega}{\omega } \cos{\left( \omega (\tau' - \tau)\right)}, \end{align}
which can be computed to arbitrary precision via Taylor series about $\tau = \tau'$,
\be G_{0,f} (\tau, \tau') 
= \frac{2}{\alpha} \ln b - \frac{\Lambda^2}{2 \alpha} \left( 1 - b^{-2} \right) (\tau - \tau')^2 + O ((\tau-\tau')^4) \ee

\subsection{General form of expectation values of exponentials}
The harmonic perturbations we consider can be re-written
\be \label{eq:SPertExp}  S_{\lambda} \left[ \theta \right] = \sum_{\pm} \frac{V_{\lambda}}{2} \int\limits_0^{\beta \hbar} d\tau ~ e^{\pm i \lambda \theta \left( \tau \right) }, \ee
and so the expectation value of $n$ copies of the perturbative terms (i.e. the sum over non-$S_0$ terms) is given by
\begin{align} \expval{S_{\lambda_1} \dots S_{\lambda_n} }_{0,f} &= \prod_{j=1}^n \left( \sum_{\lambda_j} \frac{V_{\lambda_j}}{2} \sum_{\sigma_j = \pm 1} \int\limits_0^{\beta \hbar} d\tau_j ~ \right) \expval{  \prod_{j=1}^n e^{ i \sigma_j\lambda_j \theta \left( \tau_j \right) }}_{0,f} \label{eq:SPertExp1} \\
&= \prod_{j=1}^n \left( \sum_{\lambda_j} \frac{V_{\lambda_j}}{2} \sum_{\sigma_j = \pm 1} \int\limits_0^{\beta \hbar} d\tau_j ~e^{ i \sigma_j\lambda_j \theta_s \left( \tau_j \right) } \right) \expval{  \prod_{j=1}^n e^{ i \sigma_j\lambda_j \theta_f \left( \tau_j \right) }}_{0,f}, \label{eq:SPertExp2} \\
\intertext{where the slow terms have been factored out, as they do not participate in the expectation value. This leaves}
\expval{ \prod_j e^{ i\sigma_j\lambda_j \theta_f \left( \tau_j \right) }}_{0,f} &= \expval{\exp \left\{ i\sum\limits_j  \sigma_j\lambda_j \theta_f \left( \tau_j \right) \right\} }_{0,f} \notag \\
&= \expval{\exp \left\{ i \sum\limits_j \sigma_j\lambda_j \int_f \frac{d\omega}{2\pi} \theta_f \left(\omega \right) e^{-i\omega \tau_j} \right\} }_{0,f} \label{eq:ExpExpVal1}, \\
\intertext{which is a standard Gaussian functional integral, which evaluates to}
&= \exp \left\{  \int\limits_{\Lambda/b \leq \left| \omega \right| \leq \Lambda} \frac{d\omega}{2\pi} \frac{4\pi}{\alpha \left| \omega \right|} \sum_{j,k} \left( \frac{i}{2} \sigma_j\lambda_j \right) \left( \frac{i}{2} \sigma_k \lambda_k \right) e^{i \omega \left( \tau_j - \tau_k \right)} \right\} \label{eq:ExpExpVal2} \\
&= \exp \left\{ -\frac{1}{\alpha} \int\limits_{\Lambda/b}^{\Lambda} \frac{d\omega}{\omega} \left( \sum_j \lambda_j^2 + 2 \sum_{j<k} \sigma_j \sigma_k \lambda_j \lambda_k \cos \left( \omega \left( \tau_j - \tau_k \right) \right) \right)\right\} \label{eq:ExpExpVal3} \\
&= b^{-\frac{1}{\alpha} \sum_j \lambda_j^2} \exp \left\{ - \sum_{j<k}\sigma_j \sigma_k \lambda_j \lambda_k G_{0,f} \left( \tau_j , \tau_k \right) \right\} \label{eq:ExpExpValFinal}.
\end{align}

\section{Calculating cumulants}
\subsection{Fixed part of the action: rescaling}
The cumulants are calculated with respect to the `fixed' part of the action \eqref{eq:SEff}, which dictates how we ought to rescale our coordinates and fields. We choose this to be the term generated by integrating out the bath, and require it to be self-similar (`fixed') under the RG. Thus, as we integrate out the fast modes to obtain a low energy theory, we re-scale the slow fields so that this term remains fixed as we run the RG. We have
\be \tilde{S}_0 \left[ \theta \right] = \int\limits_{-\Lambda/b}^{\Lambda/b} \frac{d\omega}{2\pi} \frac{\alpha \left| \omega \right|}{4 \pi} \left| \theta_s \left( \omega \right) \right|^2, \ee
and we can restore the original bounds of integration by introducing a rescaled frequency $\tilde{\omega} = b \omega$, giving
\be \tilde{S}_0 \left[ \theta \right] = \int\limits_{-\Lambda}^{\Lambda} \frac{d\tilde{\omega} }{2\pi b} \frac{\alpha \left| \tilde{\omega}  \right|}{4 \pi b } \left| \theta_s \left( \tilde{\omega} / b \right) \right|^2, \ee
which works out just fine if we have $\tilde{\theta} \left( \tilde{\omega} \right) = b^{-1} \theta_s \left( \omega \right)$, in which case
\be \tilde{S}_0 \left[ \theta \right] = \int\limits_{-\Lambda}^{\Lambda} \frac{d\tilde{\omega} }{2\pi } \frac{\alpha \left| \tilde{\omega}  \right|}{4 \pi } \left|\tilde{\theta} \left( \tilde{\omega} \right)\right|^2, \ee
which is, indeed, self-similar. These new frequencies suggest new temporal coordinates, $\tilde{\tau} = b^{-1} \tau$, and the scaling of $\tilde{\theta} \left( \tilde{\omega} \right)$ implies that $\tilde{\theta} \left( \tilde{\tau} \right) = \theta_s \left( \tau \right)$, preserving the form of cosine potentials.

\subsection{First cumulant: $ \expval{ \delta S_{\lambda} [\theta]}_{0,f}$}
The first order ($\mathcal{O} \left( V \right)$)contribution  to the low-energy action is given straightforwardly from \eqref{eq:SPertExp1} and \eqref{eq:ExpExpValFinal} as
\begin{align} \tilde{S}^{\left( 1 \right)}_{\vpd} &\equiv  \expval{\sum\limits_{\lambda} S_{\lambda} }~ =~ \sum\limits_{\lambda} \frac{V_{\lambda}}{2} \sum_{ \pm} \int\limits_0^{\beta \hbar} d\tau ~e^{ \pm i \lambda \theta_s \left( \tau \right) } \expval{ e^{ i \pm \lambda \theta_f \left( \tau \right) } }_{0,f} \label{eq:1stEff1} \\
&= \sum\limits_{\lambda} \frac{V_{\lambda} }{2} \sum_{ \pm} \int\limits_0^{\beta \hbar} d\tau ~e^{ \pm i \lambda \theta_s \left( \tau \right) } \left[ b^{-\frac{\lambda^2}{\alpha} } \right]  ~ = ~\sum\limits_{\lambda}   b^{-\frac{\lambda^2}{\alpha} }V_{\lambda} \int\limits_0^{\beta \hbar} d\tau \cos \left[ \lambda \theta_s \left( \tau \right) \right] , \label{eq:1stEff3} \end{align}
which we immediately recognize is self-similar, and we implement the rescaling of coordinates and fields set by the fixed part of the action $S_0$ to recover
\be \sum\limits_{\lambda}   b^{-\frac{\lambda^2}{\alpha}}V_{\lambda} \int\limits_0^{\beta \hbar/b} \left( b ~d\tilde{\tau}\right)  \cos \left[ \lambda \tilde{\theta} \left( \tilde{\tau} \right) \right] , \label{eq:1stEff4} \ee
which suggests the renormalized couplings $\tilde{V}_{\lambda} = b^{1 - \frac{\lambda^2}{\alpha}} V_{\lambda}$, and a rescaled temperature, $\tilde{\beta}\hbar = \beta \hbar/b$, and we now have full self-similarity to first order. Note that the rescaling of temperature will not affect our results. Finally, these results agree with a previous study, however in our case, the presence of a second harmonic potential requires us to examine the second cumulant.

\subsection{Second cumulant, $ -\frac{1}{2} \left\langle \left( \sum_{\lambda} S_{\lambda} - \left\langle  S_{\lambda} \right\rangle \right)^2\right\rangle_{0,f}$}
Here we consider the second cumulant, 
\be \tilde{S}^{\left( 2 \right)}_{\vpd} \equiv -\frac{1}{2} \sum_{\lambda,\lambda^{\prime}} \left( \expval{ S_{\lambda} S_{\lambda^{\prime}} } - \expval{ S_{\lambda} } \expval{ S_{\lambda^{\prime}} } \right), \ee
which we know how to calculate from \eqref{eq:ExpExpValFinal}, 
\begin{align}
\tilde{S}^{\left( 2 \right)}_{\vpd} &= - \sum_{\lambda, \lambda^{\prime}} \frac{V_{\lambda} V_{\lambda^{\prime}} }{8} \sum_{\sigma, \sigma^{\prime} = \pm 1} \int\limits_0^{\beta \hbar} d\tau  \int\limits_0^{\beta \hbar} d\tau^{\prime}  e^{ i \sigma \lambda \theta_s \left( \tau \right) } e^{ i \sigma^{\prime} \lambda^{\prime} \theta_s \left( \tau^{\prime} \right) } \notag \\
&~ \times \left\{ \expval{e^{ i \sigma \lambda \theta_f \left( \tau \right) } e^{ i \sigma^{\prime} \lambda^{\prime} \theta_f \left( \tau^{\prime} \right) } }_{0,f} - \expval{e^{ i \sigma \lambda \theta_f \left( \tau \right) } }_{0,f} \expval{ e^{ i \sigma^{\prime} \lambda^{\prime} \theta_f \left( \tau^{\prime} \right) } }_{0,f} \right\} \label{eq:2ndEff1} \\
&= - \sum_{\lambda, \lambda^{\prime}} \frac{V_{\lambda} V_{\lambda^{\prime}} }{8} \sum_{\sigma, \sigma^{\prime} = \pm 1} \int\limits_0^{\beta \hbar} d\tau  \int\limits_0^{\beta \hbar} d\tau^{\prime}  e^{ i \sigma \lambda \theta_s \left( \tau \right) } e^{ i \sigma^{\prime} \lambda^{\prime} \theta_s \left( \tau^{\prime} \right) } b^{-\frac{\lambda^2}{\alpha} } b^{- \frac{\lambda^{\prime 2}}{\alpha} } \left\{ e^{- \sigma \sigma^{\prime} \lambda \lambda^{\prime} G_{0,f} \left( \tau, \tau^{\prime} \right) } - 1  \right\} \label{eq:2ndEff2} \\
&= - \frac{1}{4} \sum_{\lambda, \lambda^{\prime}, \pm }  b^{-\frac{\lambda^2}{\alpha} }   V_{\lambda} b^{- \frac{\lambda^{\prime 2}}{\alpha} } V_{\lambda^{\prime}} \int\limits_0^{\beta \hbar} d\tau  \int\limits_0^{\beta \hbar} d\tau^{\prime}  \cos \left[  \lambda \theta_s \left( \tau \right) \pm \lambda^{\prime} \theta_s \left( \tau^{\prime} \right) \right] \left\{ e^{\mp \lambda \lambda^{\prime} G_{0,f} \left( \tau, \tau^{\prime} \right) } - 1  \right\} \label{eq:2ndEff3}.
\end{align}
We then relate this the terms allowed at bare level using a `gradient expansion' about $\tau^{\prime} = \tau$, motivated by the fact that the coarse-graining procedure should preserve temporal locality. The zeroth order term in the gradient expansion indicates that this cumulant's contribution to the RG is to generate new terms that are superpositions of the existing wave numbers. When only a single harmonic is present at bare level (i.e. the case studied by Fisher and Zwerger, or single frequency boundary sine Gordon theory), the only possibilities are to generate integer multiples of the original harmonic, which are less relevant at first order, and a correction to the kinetic term, which goes like $\omega^2$ and is thus irrelevant by power-counting. When an additional harmonic that is a non-integral multiple of the first is included, the second order processes can generate harmonics smaller and more relevant than those present at bare level. When the second harmonic is an irrational multiple of the first, these generated harmonics can be arbitrarily small.

\subsection{Two-point function contributions of higher cumulants}
As mentioned in the previous section, higher cumulants will contain products of the form 
\be \notag b^{-\frac{1}{\alpha} \sum_j \lambda_j^2 } \exp \left\{ - \sum_{j<k}\sigma_j \sigma_k \lambda_j \lambda_k G_{0,f} \left( \tau_j , \tau_k \right) \right\}  \equiv  b^{-\frac{1}{\alpha} \sum_j \lambda_j^2  }\prod_{j<k} z^{\vpd}_{j,k}, \ee 
where it will prove useful to define this quantity $z^{\vpd}_{j,k}$, which depends on $\tau_{j,k}$ and $\ell$. In general, these terms will be multiplied by factors $\exp \left( i \sigma_j \lambda_j \theta_{s} \left( \tau_j \right) \right)$ and some coefficients, and integrated over all the $\tau_j$. The re-scaling procedure will introduce factors of $b = e^{\ell}$ there as well. However, we point out a few things first.

For the second, third, and fourth cumulants, the factors that come from the two point functions will then have the respective forms
\begin{subequations} \label{eq:CumGPart}
\begin{gather}  \frac{1}{2} \left\{ 1 - z^{\vpd}_{1,2} \right\} \label{eq:2ndCumGPart} \\ 
\frac{1}{6} \left\{ 2 + z^{\vpd}_{1,2} z^{\vpd}_{1,3} z^{\vpd}_{2,3} - z^{\vpd}_{1,2}- z^{\vpd}_{1,3}  - z^{\vpd}_{2,3}  \right\} \label{eq:3rdCumGPart} \\
\frac{1}{24} \left\{ 6 - z^{\vpd}_{1,2} z^{\vpd}_{1,3}z^{\vpd}_{1,4} z^{\vpd}_{2,3}z^{\vpd}_{2,4}z^{\vpd}_{3,4}+ z^{\vpd}_{1,2} z^{\vpd}_{1,3} z^{\vpd}_{2,3} + z^{\vpd}_{1,2} z^{\vpd}_{1,4} z^{\vpd}_{2,4} + z^{\vpd}_{1,3} z^{\vpd}_{1,4} z^{\vpd}_{3,4}+ z^{\vpd}_{2,3} z^{\vpd}_{2,4} z^{\vpd}_{3,4} \right. \notag \\
\left. +  z^{\vpd}_{1,2} z^{\vpd}_{3,4} +  z^{\vpd}_{1,3} z^{\vpd}_{2,4} + z^{\vpd}_{1,4} z^{\vpd}_{2,3} - 2 \left( z^{\vpd}_{1,2}+ z^{\vpd}_{1,3}+z^{\vpd}_{1,4} +z^{\vpd}_{2,3}+z^{\vpd}_{2,4}+z^{\vpd}_{3,4} \right)  \right\} \label{eq:4thCumGPart}.
\end{gather}
\end{subequations}

These have the notable property that at $\ell = 0$, $z_{i,j}$ = 1, and the contents of the curly braces in each case sum to zero. Therefore, in taking the derivative of everything with respect to $\ell$ for the RG flow equation, only the derivatives of these cumulants \eqref{eq:CumGPart} matter (in implementing the product rule), and we take $\ell=0$ for all other terms. 

What is especially interesting is that for the third and fourth cumulants, the derivative of the factors themselves with respect to $\ell$, evaluated at $\ell = 0$, are precisely zero. Based on the way the various moments factorize for cumulants beyond the second, it appears that this will hold for all higher cumulants, and thus, to lowest order in $\Delta \ell$, there is no contribution to the RG flow equations beyond second order in the couplings $V_{\lambda}$.

\subsection{Full RG flow equation}
The full RG flow equation, to lowest order in $\ell$, is given by 
\be \frac{d}{d\ell} V_{\lambda} = \left( 1 - \frac{\lambda^2}{\alpha} \right) V_{\lambda} + \sum_{\lambda', \lambda''} C^{\lambda}_{\lambda', \lambda''} V_{\lambda'} V_{\lambda''} + \mathcal{O} \left( \ell \right) , \label{eq:RGfloweq} \ee
where $C^{\lambda}_{\lambda', \lambda''}  = \frac{\lambda' \lambda''}{2 \alpha} \left( \delta_{\lambda,\lambda'+\lambda''}- \delta_{\lambda,\abs{\lambda'+\lambda''}} \right)$, and many terms will be counted twice by the sum. 

The first order contribution to this equation, which comes from the first cumulant, is easily obtained by taking a derivative. The second order contribution follows from the standard procedure used for sine Gordon or Kosterlitz-Thouless RG: we perform a gradient expansion of \eqref{eq:2ndEff3} about $\tau'=\tau$, keeping only the lowest order term, and factor out a constant that spills over from dummy integration over $\tau'$ (i.e. in \eqref{eq:RGfloweq}, all the $V$s have been divided by this factor). In the sine Gordon case, there is only one $\lambda$, and the $\lambda + \lambda$ process generates a less relevant term, which we ignore, and the $\lambda - \lambda$ term generates a correction to the kinetic energy. In our case, we ignore the correction to the kinetic energy, as this term is already irrelevant by power coupling. Additionally, since we have two frequencies at the start of the RG, the second order processes also generate new harmonics by superposition. Hidden from \eqref{eq:RGfloweq} are terms that enter at higher order in $\ell$, including $\mathcal{O} \left( V^3 \right)$ terms, and higher order contributions from the gradient expansion
.

\section{Approximate solution of the RG flow equations}
In general, we are interested in understanding the flow of couplings when both of the bare frequencies -- which we include by hand -- are irrelevant at first order, but some linear combination of the two is \emph{relevant} for a given value of dissipation $\alpha$. We will always take the smaller of the bare frequencies to be unity, by rescaling our fields, and work with the two frequencies $\lambda = 1$, with bare coupling $u_0$, and $\lambda = \gamma$, with bare coupling $\epsilon u_0$. 

For $\gamma$ rational, it is possible to simulate the RG flow equations numerically using iterative updates, truncating at some arbitrarily large frequency, which will be highly irrelevant. However, we wish to understand what happens when $\gamma$ is \emph{irrational}, in which case there will be no means by which to truncate the allowed frequencies from below. Here we will describe a means of approximately integrating the RG flow equations in a manner that is physically sensible, in which we take $\gamma = m/n$ with $m>n$ and $m, n \in \mathbb{Z}$, and importantly, $n^{-2} < \alpha < 1$. We will take
\be \label{eq:RGintStart} \frac{dV_1}{d\ell} = \left( 1 - \frac{1}{\alpha} \right) V_1 + \dots \quad , \quad \frac{dV_{\gamma}}{d\ell} = \left( 1 - \frac{m^2}{n^2 \, \alpha} \right) V_{\gamma} + \dots , \ee
and since these couplings are guaranteed to be irrelevant, we will ignore the higher-order corrections $\dots$ to their functional form, and integrate them directly to obtain
\be \label{eq:RGintInits} V_1 \left( \ell \right) = e^{\ell \left( 1 - \frac{1}{\alpha} \right)} u_0 \quad , \quad V_{\gamma} = e^{\ell \left( 1 - \frac{m^2}{n^2 \, \alpha} \right)} \epsilon u_0 , \ee
and now for simplicity, we will take $m=n+1$ so that the relevant term $\lambda = 1/n$ is formed in a single iteration of the RG, seen from implementation of \eqref{eq:RGfloweq} :
\begin{align}   \frac{dV_{1/n} }{d\ell} & = \left( 1 - \frac{1}{n^2 \, \alpha} \right) V_{1/n} - \frac{m}{n \alpha} V_1 V_{\gamma} \notag \\
&=  \left( 1 - \frac{1}{n^2 \, \alpha} \right) V_{1/n} - \frac{m \, \epsilon u_0^2}{n \alpha} e^{\ell \left( 2 - \frac{n^2 + m^2}{n^2 \, \alpha } \right)},  \label{eq:RGintSol1}
\end{align}
and a simple Ansatz for the form of $V_{1/n}$ yields an easy solution:
\be \label{eq:RGintSol2} V_{1/n} = C \left( \ell \right) e^{\ell \left( 1 - \frac{1}{n^2\, \alpha} \right)} \Rightarrow V_{1/n} = \frac{m \, n \, \epsilon u_0^2}{n^2 \left( \alpha - 1 \right) + 1 - m^2} \left( e^{\ell \left( \frac{1}{n^2 \, \alpha} -1\right)} - e^{\ell \left( 2 - \frac{1}{\alpha}\left( 1 + \frac{m^2}{n^2}\right)\right)} \right).  \ee
We now wish to calculate $\ell^{*}$, the RG time for which $V_{1/n}$, here assumed to be the \emph{only} relevant term, grows to value unity. Since we will generally be interested in the limit where $\alpha$ is small, we will drop the latter term in \eqref{eq:RGintSol2}, and recover the expression
\be \label{eq:LstarFull} \ell^{*} = \left( 1 - \frac{1}{n^2 \, \alpha} \right)^{-1} \left( \ln \left[ \epsilon u_0^2 \right] + \ln \left[ \frac{m\, n}{n^2 \left( \alpha - 1 \right) + 1 - m^2} \right] \right), \ee
plus corrections due to the term we dropped. In general, we will be interested in the scenario where the second harmonic $V_{\gamma}$ is initially quite weak, while the main harmonic is order one (or vice versa), in which case the logarithm of $\epsilon$ will dominate, and we have
\be \label{eq:LstarLim} \ell^{*} \rightarrow \left(  \frac{1}{n^2 \, \alpha} -1 \right)^{-1} \ln \left[ \epsilon \right]  = \frac{\alpha}{\alpha_c - \alpha}  \ln \left[ \epsilon \right], \ee
where $\alpha_c$ for this rational case is $n^{-2}$, giving a straightforward scaling form for $b^{*} = e^{\ell^{*}}$ as a power of $\epsilon$.

In general, more than a single iteration of the second order RG processes will be needed to generate a relevant coupling. However, we note from the form of \eqref{eq:LstarLim} that the leading coefficient depends only on the relevant term itself, and not on any of the intermediate steps. Although this may add a somewhat complicated structure, following the procedure above will produce, at every step, additional powers of $\epsilon$ and $u_0$, factors resulting from integration as in \eqref{eq:RGintSol2}, all of which multiply a sum of terms of the form $e^{A\ell}$. All of the complications arising from integration factors can be separated into a sub-leading contribution to \eqref{eq:LstarLim}, and the only way the number of steps required to generate the relevant term shows up is in the power of $\epsilon$ (and $u_0$, if we decided to retain it). 

For an arbitrary second wave number $\gamma$, it would be quite cumbersome to determine the optimal sequence of second order processes to generate a given, relevant term. However, for the particular choice of $\gamma = \varphi = \frac{1}{2}\left( 1 + \sqrt{5} \right)$, this is quite simple. For a given value of $\alpha$, the relevant term generated in the fewest number of RG iterations will be a Fibonacci wave number, the $n^{\rm th}$ of which is given by $\lambda_n \equiv \left( -1 \right)^n \left( F_{n+1} - \varphi F_n \right) = \varphi^{-n}$. In the first step, the term $\lambda_1 = \varphi^{-1} = \varphi -1$ is generated, and in general, $\lambda_n$ is formed in $n$ `steps'. Correspondingly, after $n$ steps, there are $F_n$ factors of $\epsilon$, where $F_n$ is the $n^{\rm th}$ Fibonacci number, and $F_{n+2}$ factors of $u_0$. Thus, supposing $n^{*}$ corresponds to the first Fibonacci potential with a coupling that grows to $O(1)$, the RG time required to do so is given by
\be \label{eq:LstarFib} \ell^{*} \to \ell^{\vps}_{n^*} = \frac{\alpha \, F_{n^{*}}}{\varphi^{-2n^{*}} -\alpha}  \ln \left[ \epsilon \right] , \ee
plus sub-leading corrections. In the limit of large $n^{*}$, the parenthetical pre-factor goes quickly to $- F_{n^{*}}$. For scenarios in which the first relevant $\lambda_{n}$ is just barely relevant, it may be the case that $\lambda_{n+1}$ or a subsequent Fibonacci potential grows to $O(1)$ faster than $\lambda_n$, despite being formed later in the RG, and thus we must always choose the minimal value of $\ell^{*}$ as the minimal $\ell_n$ among the relevant $\lambda_n$, for a given value of $\alpha$.

\section{Localization length}
While the RG time $\ell^{*}$ provides a useful cross-over scale for when the particle begins to `feel' localized due to the presence of a growing, relevant potential, we also note that it can take a substantial amount of RG time for this localization to take hold. Although the RG flow equations clearly show that, for any non-zero value of $\alpha$, there will exist some potential  generated by second order processes that is relevant, it is also clear that as $\alpha$ is decreased, it will take longer and longer for this localization to take effect, and additionally, the potential well to which the particle is localized will be correspondingly larger. To give this physical meaning, we will extract a localization length from the crossover time, $\ell^{*}$. 

Although it is possible to associate a genuine time scale $\tau$ to the RG time $\ell^{*}$, and then calculate how far the particle travels in such a time scale, the same result is more expediently recovered using the following definition:
\be \xi^* = \expval{q^2 \left( \tau \right)}^{1/2} = \frac{q_0}{2\pi} \sqrt{ \expval{\theta^2 \left( \tau \right)}}, \ee
and we perform the average by integrating out only the modes $\Lambda/b^* \leq \omega \leq \Lambda$, i.e. those corresponding to our RG time $\ell^*$. Additionally, we make the assumption that, until the time $\ell^{*}$ is reached, the effect of the cosine terms may be ignored in evaluating the two-point function, and we have
\be \expval{\theta^2 \left( \tau \right)} = 2 \int\limits_{\Lambda/b^*}^{\Lambda} \frac{d\omega}{2\pi} \left[ \frac{\alpha \omega}{2 \pi}  + \dots \right]^{-1} 
= \frac{2}{\alpha} \left. \ln \left( \omega  \right) \right|^{\Lambda}_{\Lambda/b^*} = \frac{2}{\alpha} \ln b^* = \frac{2 \ell^*}{\alpha}, \ee
and from this we conclude that
\be \xi^* =  \frac{q_0}{2\pi} \sqrt{\frac{2 \ell^*}{\alpha}}.\ee
We can compare to a previous prediction that the localization length diverges as $\left( \alpha - 1 \right)^{-1/2}$ for the case of integer harmonics by using the result for \eqref{eq:LstarLim} for $n =1$, which indeed gives $\xi^* \sim \left( \alpha - 1 \right)^{-1/2}$. 

\end{appendix}

\end{document}